\documentclass[useAMS,usenatbib]{mn2e}
\usepackage{amsmath}
\usepackage{graphicx}
\newcommand{\ms}{\rm {h^{-1}M_{\odot}}}
\newcommand{\mpch}{{\rm h^{-1}Mpc}}

\begin{document}
\bibliographystyle{mn2e}
\graphicspath{{fig1/}}
\title[]{Constraining the substructure of dark matter haloes with galaxy-galaxy lensing}
\author[Ran Li et. al]
       {\parbox[t]{\textwidth}{
        Ran Li$^{1}$\thanks{E-mail:ranl@bao.ac.cn},
        H.J. Mo$^{2}$,
        Zuhui Fan$^{3}$,
        Xiaohu Yang$^{4}$,
         Frank C. van den Bosch$^{5}$}
        \vspace*{3pt} \\
  $^{1}$National Astronomy Observatory, Chinese Academy of Sciences, Beijing 100871, China\\
  $^{2}$Department of Astronomy,  University of Massachusetts, Amherst
        MA 01003, USA \\
  $^{3}$Department of Astronomy, Peking University, Beijing 100871, China\\
  $^{4}$Shanghai Astronomical Observatory, the Partner Group of MPA,
        Nandan Road 80, Shanghai 200030, China\\
  $^{5}$Astronomy Department, Yale University, P.O. Box 208101
        New Haven, CT 06520-8101, USA}

\maketitle

\begin{abstract}
  With galaxy groups constructed from the Sloan Digital Sky Survey
  (SDSS), we analyze the expected galaxy-galaxy lensing signals around
  satellite galaxies residing in different host haloes and located at
  different halo-centric distances. We use Markov Chain Monte Carlo
  (MCMC) method to explore the potential constraints on the mass and
  density profile of subhaloes associated with satellite galaxies from
  SDSS-like surveys and surveys similar to the Large Synoptic Survey
  Telescope (LSST).  Our results show that for SDSS-like surveys, we
  can only set a loose constraint on the mean mass of subhaloes.  With
  LSST-like surveys, however, both the mean mass and the density
  profile of subhaloes can be well constrained.
\end{abstract}



\section{introduction}

In the cold dark matter (CDM) scenario, large-scale structures in the
universe grow hierarchically through gravitational
instabilities. Galaxies are assumed to form in dark matter potential
wells through gas cooling and star formation
\citep{white1978,white1991}. During the hierarchical formation
process, when small haloes merge into larger systems, they become
subhaloes.  High resolution simulations show that while some of them
are disrupted due to processes such as tidal stripping and impulsive
heating, a large fraction of the subhaloes survive. Hence, probing
the masses and density profiles of the population of subhaloes is
a key test for the CDM structure formation model.

The mass function, spatial distribution and density profile of
subhaloes has been extensively studied with semi-analytical models as
well as $N$-body simulations \citep[e.g.,][]{Hayashi2003, gao2004,
  Bosch05, giocoli08, Giocoli10, springel08_aqu, zentner2003,taylor2004,
  oguri2004,gill2004}. State-of-the-art, high-resolution simulations
\citep{springel08_aqu,diemand2007} can resolve subhaloes down to a mass
of $\sim 10^7\ms$, and thus provide detailed predictions for both
their mass function and density profiles.
On the other hand, it is very challenging to probe dark matter
subhaloes observationally because of their darkness and the relatively
weak gravitational potential compared to that of their host haloes.
Arguably the best (and most direct) probe of dark matter substructure
is gravitational lensing.  The existence of substructure in a smooth
dark matter halo induces flux-ratio anomalies for multiple images of a
lensed system \citep{Mao1998, Metcalf2001, Mao2004, Kochanek2004,
  Maccio2006, Xu2009}, and also perturbs the surface brightness of
extended Einstein rings and arcs \citep{Koopmans2005, Vegetti2009a,
  Vegetti2009b, Vegetti2010, Vegetti2012}. So far, about 200
galaxy-sized strong lensing systems have been discovered
\citep[e.g.,][]{bolton2008}.  The constraints on the mass fraction of
subhaloes in galaxies have been investigated \citep[e.g.,][]{Xu2009}.
However, the number of strong lensing systems with high quality
imaging observations is still limited.  Furthermore, strong lensing
effects can only probe the very central region of galaxies. Therefore
it is not easy to obtain a general understanding about subhaloes from
strong lensing effects alone.

Because most of the satellite galaxies are thought to reside in
subhaloes, galaxy-galaxy lensing can be an effective way to probe
subhaloes statistically. While it was first developed to estimate the
dark matter distribution of massive systems, the recent advance of
wide and deep surveys, such as the Sloan Digital Sky Survey
(SDSS)\footnote{http://www.sdss.org} 
and the Canada-France-Hawaii Telescope Legacy Survey
(CFHTLS)\footnote{http://www.cfht.hawaii.edu/Science/CFHLS/}, has
allowed the application of galaxy-galaxy lensing analyses to the 
study of the mass distribution around lens galaxies of different luminosities,
stellar masses, colors, and morphological types
\citep[e.g.,][]{Brainerd1996, Hudson1998, Hoekstra2003, Hoekstra2004,
  Mckay2001, Mandelbaum2005, mandelbaum2006, Mandelbaum2008,
  sheldon2009, jonhston2007}. Several studies have investigated the
potential of using galaxy-galaxy lensing to probe the masses and
density profiles of dark matter subhaloes \citep[e.g.,][]{Yang2006,
  Li2009, Mira2011}.  Current observations can only set partial
constraints on subhalo properties for individual massive clusters of
galaxies \citep[e.g.,][]{Limousin2007,Natarajan2007,Natarajan2009}.
However, with the next generation of large surveys, such as the Large
Synoptic Survey Telescope (LSST) \footnote{http://www.lsst.org/}, the
surface number density of source galaxies that can be used for
galaxy-galaxy lensing analyses can reach $n_g\sim 50\hbox{
  arcmin}^{-2}$, in comparison with $n_g\sim 1\hbox{ arcmin}^{-2}$ for
SDSS. This will significantly increase the signal-to-noise of the
lensing signal of dark matter subhaloes, thus enabling direct
measurements of their masses and density profiles.

The goal of this paper is to examine the potential of using
galaxy-galaxy lensing to constrain the properties of dark matter
subhaloes, such as their mass and density profile. The subhalo
properties of satellite galaxies with some fixed property (i.e.,
stellar mass) are likely to depend on both the host halo mass and the
location of the satellite galaxy within the host halo. In order 
to probe these dependencies, we need to distinguish satellite
galaxies located in different haloes and at different distances from the
centers of their host haloes.  One way to do this is to select lens
galaxies using a galaxy group catalog \citep{Yang2006, jonhston2007,
  sheldon2009, Li2009}. This allows one to select as lenses a subset
of satellite galaxies that reside in haloes (groups) of similar mass,
and that are at similar (projected) distances from their halo (group)
center.  In \cite{Li2009}, we applied such a method to predict
galaxy-galaxy lensing effects for lens galaxies of different
luminosities and different morphological types using a group catalog
constructed by \citet[][hereafter Y07]{Yang2007} from the SDSS.  The
predictions are found to agree well with lensing observations of SDSS from
\citet{mandelbaum2006}, 
 demonstrating the validity of the method.  In this paper, we
use the same methodology to predict the galaxy-galaxy lensing effects
for satellite galaxies selected from the SDSS-DR7 group catalog.  We
investigate the corresponding signal detectability with current and
next generation surveys. Employing a Markov Chain Monte Carlo (MCMC)
method, we further explore the possibility of constraining both the
subhalo and host halo properties in lensing observations at different
noise levels expected from different surveys.

This paper is organized as follows. We provide a brief description of
the galaxy-galaxy lensing basics in Section \ref{sec:gglensing}, and
discuss the modeling method in Section \ref{sec:model}.  In Section
\ref{sec:sdssgc}, we introduce the group catalog, SDSSGC, from which
lens galaxies are selected.  In Sections \ref{sec:host} and
\ref{sec:subhalo}, we describe our models for the dark matter
distribution around galaxies. In Sections \ref{sec:data} and
\ref{sec:mcmc}, we show the results and examine the detectability of
the predicted lensing signals in SDSS-like and LSST-like surveys. We
discuss some systematic bias in our method, and how to correct for it,
in Section \ref{sec:discussion}. Section \ref{sec:summary} contains a
summary.

Throughout the paper, we adopt a $\Lambda$CDM cosmology 
with parameters given by the WMAP-7-year data \citep{wmap7}.

\section{Galaxy-galaxy lensing}
\label{sec:gglensing}

Galaxy-galaxy lensing measures the tangential shear, $\gamma_t(R)$,
azimuthally averaged over a thin annulus at the projected radius $R$
around the lens galaxies. In the weak lensing regime, this quantity is
related to the excess surface density, $\Delta\Sigma$ (hereafter ESD)
through the relation
\begin{equation}\label{eq:ggl}
\Delta\Sigma(R)=\gamma_t(R)\Sigma_{\rm crit}=\bar{\Sigma}(<R)-\Sigma(R)\,,
\end{equation}
where $\bar{\Sigma}(<R)$ is the average surface mass density within
$R$, and $\Sigma(R)$ is the azimuthally averaged surface density at
$R$. It is noted that there is no mass-sheet degeneracy here, and
$\Delta\Sigma(R)$ is independent of a uniform background. In the above
equation,
\begin{equation}
\Sigma_{\rm crit}=\frac{c^2}{4\pi G}\frac{D_s}{D_l D_{ls}(1+z_l)^2}
\end{equation}
is the critical surface density in comoving units, with
$D_s$ and $D_l$ the angular diameter distances to the lens and to the source,
$D_{ls}$ the angular diameter distance between the lens and the source, 
and $z_l$ the redshift of the lens.

The lensing signal around a galaxy is determined by the projected
density profile around it. On average, the surface mass density,
$\Sigma(R)$, is related to the line-of-sight projection of the
galaxy-matter cross-correlation function, $\xi_{\rm g, m}(r)$. Under
the approximation that lenses are at distances much larger than $R$,
we can write
\begin{equation}\label{xi_gm}
\Sigma(R)=\bar{\rho}\int 
\left[1+\xi_{\rm g,m}(\sqrt{R^2+\chi^2}) \right ]\,d\chi\,;
\end{equation}
and
\begin{equation}\label{xi_gm_in}
\Sigma(< R)=\frac{2}{R^2} \int_0^{R} \Sigma(u) u du ,
\end{equation}
where $\bar{\rho}$ is the mean density of the universe 
and $\chi$ is the comoving radial distance along the line of sight.  

The lensing signal around a satellite galaxy depends sensitively on its
location in the host dark matter halo \citep{Yang2006, Li2009}.  The
$\Sigma(R)$ around a central galaxy, which mostly resides at the
center of the host dark matter halo, is dominated by the density
profile of its host halo. On the other hand, the lensing signal of a
satellite galaxy, which orbits in the host halo, consist of two
parts. On small scales, the signal is dominated by the subhalo
associated with the satellite itself.  On larger scales, however, the
lensing signal is mainly due to the host halo. We therefore need to
model the density profiles of both host haloes and subhaloes.

When calculating the surface mass density around a satellite, we
neglect the contributions from other subhaloes.  This approximation
is not expected to lead to large errors because the fraction of mass contained
in subhaloes is only about 10\% of the total mass of the host halo
\citep[e.g.][]{Bosch05, springel08_aqu, Giocoli10}. For a single
halo, these subhaloes produce small fluctuations on the host halo
profile.  However, in galaxy-galaxy lensing analysis where one stacks the
lensing signal around many lens galaxies, the net contribution
from subhaloes other than the ones associated with the satellite
galaxies themselves is averaged out and included in the host halo
profile under the assumption that subhalos in a host halo are
not correlated. In the model calculation, we also neglect the two-halo term,
i.e. the contribution to the lensing signal from other haloes
in the foreground and background.  Our previous studies \citep{Li2009,
  Cacciato2009} calculated this contribution with different methods,
and showed that the two-halo term is completely negligible on the scales
of individual haloes we are concerned with here.

\section{Modeling the structure of dark matter haloes}
 \label{sec:model}
 
In this paper, we adopt the same methodology as that used in \cite
{Li2009} to model the galaxy-galaxy lensing signal around a sample of
satellite galaxies. In the following subsections, we describe briefly
the galaxy and group catalogs, and our models for the mass
distributions around satellite galaxies.

\subsection{Galaxy groups}
\label{sec:sdssgc}

In \citet{Li2009}, we used the SDSS DR4 group catalog
\citep{Yang2007}.  Here we use an updated version of 
this catalog\footnote{http://gax.shao.ac.cn/data/Group.html} (hereafter SDSSGC)
based on the SDSS DR7 \citep{Abazajian2009}.  The group catalog is
constructed with the adaptive halo-based group finder developed by
\citet{Yang2005,Yang2007} using galaxies with spectroscopic redshifts
in the range of $0.02 \le z \le 0.2$.  The redshift completeness is
$\mathcal{C} > 0.7$. Three group samples with different sources of
galaxy redshifts have been constructed.  Our analysis is based on
Sample II, which consists of $599301$ galaxies with redshift from the
SDSS and $3269$ galaxies with redshift from other sources.  There are
in total $472113$ groups, including those with only one member galaxy.
\footnote{Following Y07, we refer to a system of galaxies as a group regardless of its richness and mass,
including isolated field galaxies (i.e., groups with one member) and clusters of galaxies.}

A key aspect of this group finder is to estimate the halo mass, $M$,
for each group with a ranking method. In SDSSGC, two estimators for
halo mass are provided. One is based on the characteristic luminosity
of a group, defined to be the total luminosity of all member galaxies
with $M_r - 5 \log{h} < -19.5$. The other is based on the
characteristic stellar mass, $M_{\rm stellar}$, defined to be the
total stellar mass of members galaxies with $M_r - 5 \log{h} <
-19.5$. The stellar mass of an individual galaxy is calculated from its
luminosity and colors using the fitting formula given by
\citet{Bell2003}. Y07 showed that the characteristic stellar mass is a
better indicator of the halo mass, and thus we adopt this mass
estimator throughout the paper.

The basic assumption of the ranking method to assign a halo mass to a
group is that there is a one-to-one relation between $M_{\rm stellar}$
and the halo mass. Once a theoretical dark matter halo mass function
is adopted, one can establish a relation between halo mass, $M$, and
$M_{\rm stellar}$ so that the number of haloes with masses above $M$ is
equal to the number of groups with characteristic stellar mass above
$M_{\rm stellar}$. A group with a given $M_{\rm stellar}$ is then
assigned the corresponding halo mass $M$. Clearly, this one-to-one
mapping requires the group sample to be complete. Therefore, we only
use complete samples of groups in the SDSSGC in our ranking. The
masses of other groups are estimated using linear interpolation based
on the $M_{\rm stellar}$ -$M$ relation obtained from the complete
sample. We refer readers to \citet{Yang2007,Yang2008} for details
about the group catalog construction and the halo mass
assignment. According to Y07, the uncertainty in the mass assignment
is about 0.2-0.3 dex for groups considered in this paper. This
uncertainty will not change our results significantly because the
considered lensing signals are the average signals over a statistical
sample of galaxies. We have tested the effect by performing
calculations with group mass to which an artificial 0.3 dex log-normal
error is added. This uncertainty brings negligible change in our
results.  Note that the halo mass assigned to a group in the SDSSGC is
$M_{200}$, which is the mass enclosed in the radius, $r_{\rm 200}$,
defined such that $M_{\rm 200} = 4\pi r^3_{\rm 200}(200\bar{\rho})/3$.
For consistency, we convert $M_{\rm 200}$ to our definition of halo
mass (see Eq.~[\ref{eq:Mass}] below) using the conversion method
described in the appendix of \citet{Hu2003}.
 
\subsection{Host halo density profile}
\label{sec:host}

We assume that the host dark matter halo of each group is centered on
the most massive group member, to which we refer as the central
galaxy. The dark matter host haloes are assumed to follow the NFW
\citep{NFW97} profile,
\begin{displaymath}
  \rho_{\rm dm}(x)=\left \{ \begin{array}{ll}
    \frac{\delta_0{\rho_{\rm crit}}}{x(1+x)^2} & \textrm{if \it x $\leq$ c} \\
    0 & \textrm{otherwise}
    \end{array} \right . \,,
  \label{eq_NFW}
\end{displaymath}
where $x \equiv r/r_s $, with $r_s$ being the characteristic scale of
the halo and related to the halo virial radius $r_{\rm vir}$ through
the concentration parameter $c=r_{\rm vir}/r_s$, and $\rho_{\rm crit}$
is the critical density of the universe. The characteristic
over-density $\delta_0$ is related to the average over-density of a
virialized halo, $\Delta_{\rm vir}$, by
\begin{equation}
\delta_0={\Delta_{\rm vir}\over 3}{c^3\over \ln(1+c) -c/(1+c)}\,.
\end{equation}
For the $\Lambda$CDM model considered here, we adopt the parametric
form of $\Delta_{\rm vir}$ given by Bryan \& Norman (1998) based on
the spherical collapse model,
\begin{equation}
\Delta_{\rm vir}=18\pi^2+82[\Omega(z)-1]-39[\Omega(z)-1]^2\,,
\end{equation}
where $\Omega(z)$ is the cosmological density parameter at redshift
$z$. The viral mass of a halo can then be written as:
\begin{equation}
\label{eq:Mass}
M = {4\pi\over 3} r^3_{\rm vir}\Delta_{\rm vir} \rho_{\rm crit}\,.
\end{equation}
It is clear that for a given halo mass, the halo density profile depends
only on the concentration parameter $c$.  Numerical simulations show
that at a given redshift, $c$ decreases gradually with halo mass
\citep[e.g.,][]{Bul01,Eke01}. However the exact mass dependence of the
concentration parameter has not yet been well constrained by
observations. Various fitting formulae have been proposed on the bases
of numerical simulations \citep[e.g.,][]{Bul01, Zhao03, Dolag04,Maccio07,
Zhao09}. Since the difference between these different
fitting functions does not affect our results qualitatively, we adopt
the fitting formula of \cite{Bul01},
\begin{equation}
\label{eq:concentration}
c={c_{\ast}\over 1+z}\left({M\over 10^{14}\ms}\right)^{-0.13} \,,
\end{equation}
where $c_{\ast}\approx 8$ for the $\Lambda$CDM cosmology considered
here. Although simulations indicate a scatter of $\sim 0.1$dex in this
concentration - mass relation \citep[e.g.][]{Jing00, Bul01, Wechsler02,
  Maccio07}, we ignore this scatter in our analysis.  We have verified
that adding scatter has no significant impact on any of our results.

To obtain galaxy-galaxy lensing signals around a halo center, we need
to project the 3-D mass distribution.  According to
\citet{Hamana2004}, the ESD can be written as
\begin{equation}
\Delta\Sigma(y)=2\Delta\Sigma_s f(y)\,, 
~~~~~~
y=\frac{R}{r_s} \,,
\end{equation}
where $\Delta\Sigma_s=\rho_s r_s $. For the NFW profile, the
dimensionless function $f(y)$ can be written as
\citep{Wright00,Bartelmann96}
\begin{equation}
f(y)=\left\{
\begin{array}{ll}
  {1\over (y^2-1)}\left(1 - {\ln\left( \frac{1+\sqrt{1-y^2}}{y}\right)
       \over\sqrt{1-y^2}}
	 \right) & \mbox{if $y < 1$ ;} \\
      \frac{1}{3} & \mbox{if $y=1$; } \\
      {1\over (y^2-1)}\left(1 - { {\rm atan}{\sqrt{y^2-1}}\over \sqrt{y^2-1}}
	\right)  & \mbox{if $y > 1$ .} 
\end{array} \right.\,
\end{equation}
%
Note that the above equation assumes that the NFW profile extends
infinitely without a truncation at the virial radius. We have tested
that the difference between truncated and non-truncated profiles has
negligible impact on our results.

\subsection{Subhalo density profile and mass distribution}
\label{sec:subhalo}

In order to assign a subhalo mass to a satellite galaxy we use the
evolved subhalo mass function of \citet{Bosch05}, which describes the
abundance of subhaloes as function of their evolved, present-day mass.
Specifically, for each satellite galaxy, we assume that the mass of
its subhalo at accretion (i.e., its un-evolved mass) is related
monotonically to its stellar mass. The retained mass fraction of the
subhalo after evolution in the host halo can be described by a
parameter $f_m$.  \citet{gao2004} analyzed the radial dependence of
$f_m$ from a large sample of subhaloes in cosmological simulations,
and showed that
\begin{equation}\label{eq_fm}
f_m=0.65 (r_{\rm dis}/r_{\rm vir})^{2/3}\,,
\end{equation}
where $r_{\rm dis}$ is the distance of the subhalo from the center of
the host halo, and $r_{\rm vir}$ is the virial radius of the host
halo. Observationally, the line-of-sight distance cannot be estimated
accurately, and only the projected halo-centric distance, $r_{\rm p}$,
can be used. We then adopt the following approach to estimate the
three-dimensional distance of the satellite. For a satellite with a
given $r_{\rm p}$, we randomly sample a 3-D halo-centric distance
assuming that the spatial distribution of the satellites follows the
NFW profile, and use this distance in equation (\ref{eq_fm}) to
estimate $f_m$. With $f_m$ obtained for each satellite galaxy,
we can define a ranking parameter
$\cal{Q}$ for every member satellites in a group with host halo mass
$M$ as
\begin{equation}
{\cal Q}=f_m M_{\rm \ast }\,,
\end{equation}
where $M_{\rm \ast }$ is the stellar mass of the satellite galaxy.  We
then generate a set of subhalo masses for a given host halo mass $M$
using the fitting formula for the evolved subhalo mass function of
\citet{Bosch05}.  Finally, by a mapping between the ranks 
in subhalo masses and in the value of ${\cal Q}$,  a subhalo mass is
assigned to a satellite according to its ranking parameter $\cal{Q}$.
\begin{figure*}
\begin{center}
\includegraphics[width=0.9\textwidth]{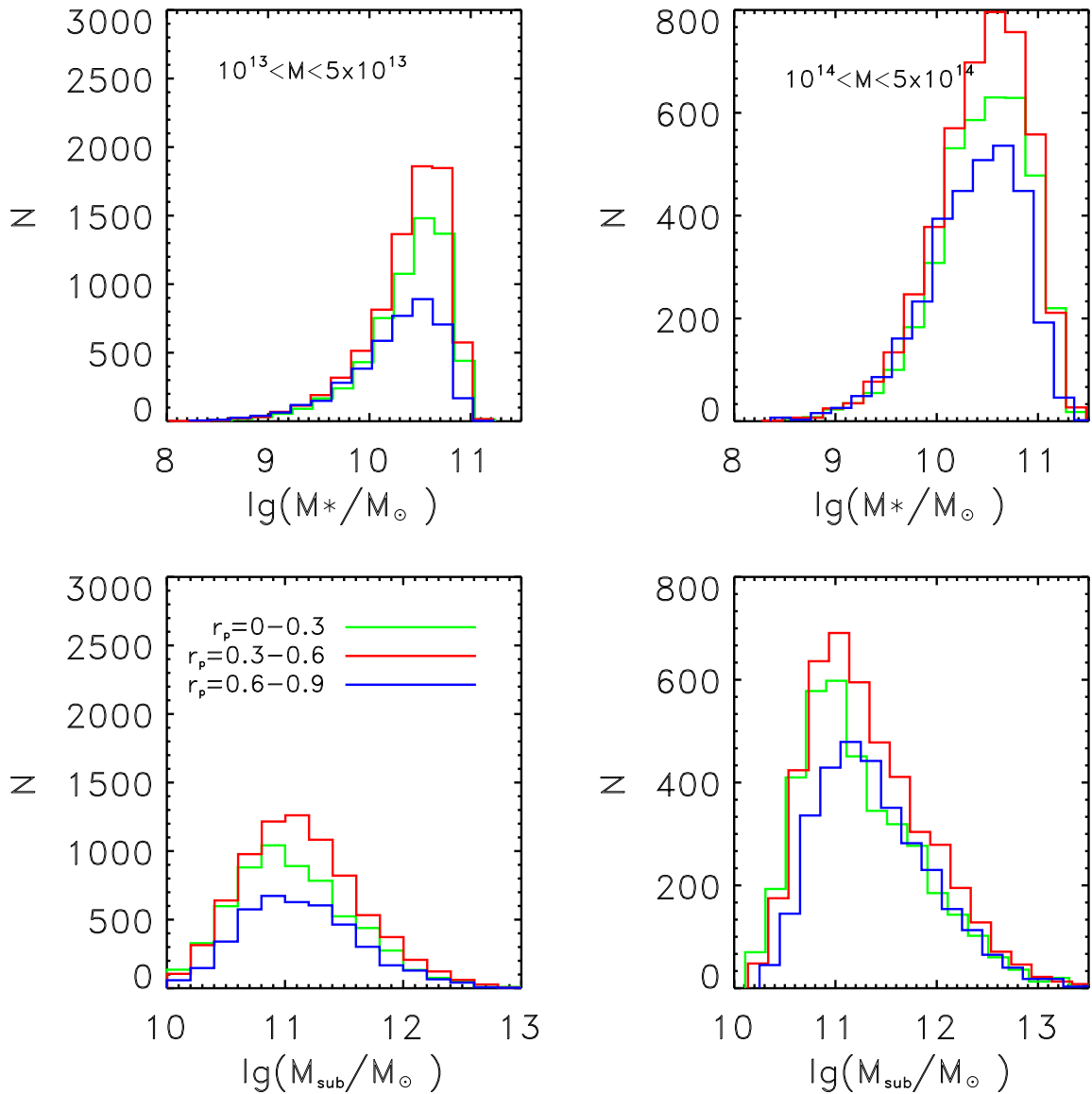}
\end{center}
\caption{The figure shows the stellar mass and subhalo mass
  distribution of satellite galaxies with host halo mass in certain
  ranges. The host halo mass ranges are $[10^{13},5\times10^{13}]\ms$
  for left panels and $[10^{14},5\times10^{14}]\ms$ for right
  panels. In the upper panels of the figure, we plot the stellar mass
  distribution for the satellites.  We split the satellites into
  sub-samples according to their projected halo-centric distances
  $r_{\rm p}$ and plot the stellar mass distributions of these
  sub-samples with different colors. The $r_{\rm p}$ ranges are
  marked in bottom left panel in unit of $r_{\rm vir}$. 
  In the lower panels, we plot the mass distribution
  of the subhaloes for the corresponding satellite sample.}
\label{fig:msdis}
\end{figure*}

Fig.\ref{fig:msdis} shows the results of our subhalo mass assignment.
The upper panels are the stellar mass distributions for the satellites
with their host halo mass in the range indicated in the plots.  We
split satellites into sub-samples according to their projected
halo-centric distance $r_p$ shown by different lines.  One can find
that the peak of the stellar mass distribution does not depend
strongly on their projected distance. The mass distributions of the
corresponding subhaloes are shown in the lower panels. The average 
dark matter mass to stellar mass ratio is about $10$.

We model the density profile of subhaloes with a truncated NFW profile 
\begin{equation}\label{eq:rhosub} 
\rho_{sub}(r)=\left\{
\begin{array}{rl} f_t \rho_{i,sub}(r) & \text{if } r \le r_t,\\
    0 & \text{if } r > r_t
\end{array} \right. 
\end{equation}
where $\rho_{i,sub}(r)$ is the NFW profile corresponding to the mass
of the subhalo at the time of its accretion into its host.  The
parameter $f_t$ is a dimensionless factor describing the reduction in
the central density, and $r_{t}$ is a cut-off radius imposed by the
tidal force of the host halo. In $\rho_{i,sub}(r)$, the characteristic
scale and density are denoted by $r_{\rm s,sub}$ and $\delta_{0,sub}$,
respectively. Note that the parameters of $f_t$ and $\delta_{0,sub}$
can be combined to a single parameter denoted by $\rho_{\rm 0,sub}$.
For $f_t=1$ and $r_{t}\gg r_{\rm s,sub}$, $\rho_{sub}(r)$ approaches
to the standard NFW profile $\rho_{i,sub}(r)$.
For the cut-off radius $r_{t}$, we use the analytical tidal 
radius formula \citep{BT87,Tormen98},
\begin{equation}\label{eq:rt}
r_t=\left( \frac{M_{\rm sub}}
{(2-d\ln M/d\ln r)M(<r_{\rm dis})}  \right)^{1/3} r_{\rm dis}\,,
\end{equation}
where $M(<r_{\rm dis})$ is the host halo mass within a sphere of
radius $r_{\rm dis}$. As shown by \citet{springel08_aqu}, this
analytical prediction agrees well with the trunctation radii of dark
matter subhaloes in $N$-body simulations.  The density profile is
normalized to the mass assigned to the subhalo by choosing a proper
$f_t$ (or equivalently $\rho_{\rm 0,sub}$) . Therefore in our model
the mass profile assigned to a subhalo is specified by three
quantities: (i) the stellar mass of the satellite galaxy; (ii) the
host halo mass; and (iii) the distance between the satellite and the
center of the host halo.

It should be pointed out that there are still substantial
uncertainties in modeling the mass distribution around individual
satellite galaxies. In particular, many of the results about subhaloes
are obtained from pure $N$-body simulations.  It is unclear how
significant the effect of including baryonic matter is. However, the
aim of this work is to investigate to what extent current and future
lensing data can constrain the density distribution of dark matter
subhaloes.  Our relatively simple model for the density distribution of
subhaloes should be sufficient for this purpose.

\section{Modelled lensing signal} 
\label{sec:data}
\subsection{Lensing signal of individual satellite}
\label{sec:single}

\begin{figure*}
\includegraphics[width=\textwidth]{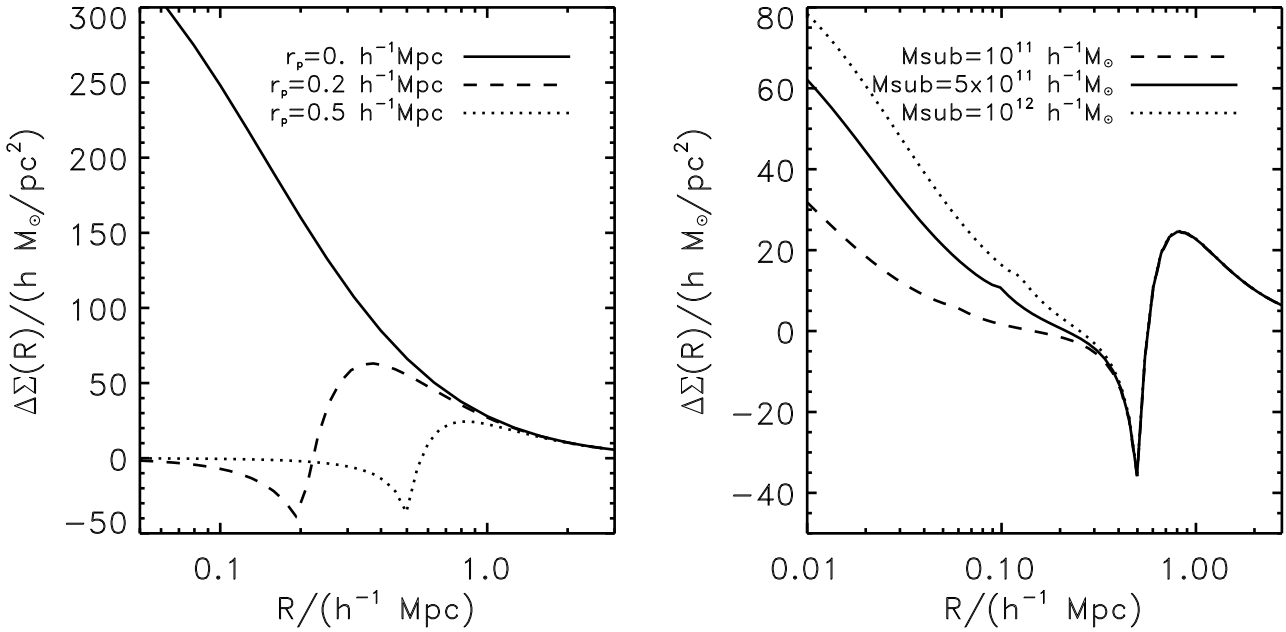}
\caption{Left panel: host halo contribution to lensing signal
 around satellites at different positions. The host halo mass is set to $10^{14}\ms$.
 For clarity, we only plot the contribution of host halo and omit the 
subhalo contribution.  Different lines represent different projected
halo-centric distances.  Right panel: different lines show
lensing signals around satellites of different masses in
a host halo of $10^{14}\ms$ at halo-centric distance of 0.5 $\mpch$. }
\label{fig:model}
\end{figure*}

We first calculate the behavior of galaxy-galaxy lensing signal as a
function of the projected radius around a single satellite.  Similar
results can be found in, e.g. \citet{Yang2006} and \citet{Li2009}. In
Fig.\ref{fig:model}, we plot $\Delta\Sigma(R)$ for satellites of
different mass and at different position in a host halo with mass of
$M=10^{14}\ms$. In the left panel, the subhalo mass is set to be zero
to show the lensing signal contributed by the host halo
alone. Different lines show the predictions for satellites located at
different projected halo-centric distances, $r_p$. It is clear that
the contribution from the host halo depends strongly on the position
of the satellite. For $r_p=0$, i.e., the central galaxy, $\Sigma(R)$
is just the projection of NFW density profile, and $\Delta\Sigma(R)$
decreases monotonously.  For a satellite galaxy with $r_p\neq 0$,
however, $\Delta\Sigma(R)$ from the host halo is nearly 0 on small
scales around the satellite. This is because the host halo density
varies smoothly on small scale around the satellite. As $R$ grows,
$\Delta\Sigma(R)$ decreases to negative values, reaching a minimum at
$R=r_p$ where the outer annulus in Eq.(\ref{eq:ggl}) reaches the
centre of the host halo. It then goes up rapidly, eventually
approaching the $\Delta\Sigma(R)$ profile for the central galaxy.  In
the right panel, we show $\Delta\Sigma(R)$ for satellites at
$r_p=0.5\mpch$ with different subhalo mass. Clearly subhalo dominates
the inner part of the ESD profile.  The value of $\Delta\Sigma(R=0.01
\mpch)$ increases by a factor of $2.5$ when the subhalo mass increases
from $10^{11} \ms$ to $10^{12} \ms$.

\subsection{Stacking}

Observationally, weak lensing signals are derived by accurately
measuring the shape of the light distribution of source galaxies.  The
intrinsic shape of source galaxies then contributes significant noise
to the lensing signal.  Specifically, the measured tangential
ellipticity $e_+$ of a source galaxy is related to the lensing
tangential shear $\gamma_t$ acting upon it by
\begin{equation}
  e_+=2\gamma_t\mathcal{R}+e^{int}_+ \,,
  \label{shear}
\end{equation}
where $e^{int}_+$ is the intrinsic tangential ellipticity of the
source galaxy and $\mathcal{R}$ is the ``responsivity'', reflecting
how the shape of a galaxy responds to the shear applied to it
\citep{B_J2002}. This quantity can be determined from observational
data, and we set $\mathcal{R}=0.87$ following \citet{mandelbaum2006}.
In galaxy-galaxy lensing, one suppresses the noise arising from the
intrinsic shape of source galaxies by combining shape information from
as many source images as possible. Unless the density of background
source is extremely high, this typically requires also stacking the
signal from many lens galaxies. Assuming no intrinsic alignment for
source galaxies, the averaged intrinsic ellipticity over many galaxies
is expected to approach zero.  Thus the average of $e_+$ gives rise to
an unbiased estimate of $2\gamma_t\mathcal{R}$. The corresponding
uncertainty of the tangential shear measurement can be written as
\begin{equation}\label{eq:noise}
  2\mathcal{R}\sigma_{\gamma}=\sqrt{\sigma_{SN}^2+\sigma_e^2}
  /\sqrt{N_{pair}} \,,
\end{equation}
where $N_{pair}$ is the number of lens-source pairs,
$\sigma_{SN}\approx0.3$ is the source galaxy intrinsic shape
dispersion for one component of the ellipticity, and $\sigma_e$ is the
measurement noise for one component of the ellipticity.  The
measurement noise can originate from photon noise and inaccurate PSF
corrections.  For the SDSS, $\sigma_e$ falls in the range from $0.05$
to $0.4$, depending on the luminosity of the source galaxies
\citep{Mandelbaum2005}. Throughout this paper, we adopt
$\sigma_e=0.2$. It should be pointed out that the LSST will improve
significantly in survey depth and angular resolution compared to SDSS,
and thus $\sigma_e$ is expected to be much smaller. The measurement
noise adopted here is therefore a very conservative estimate for
future surveys.  We fix the lens redshift to be $z_l=0.15$, which is
the mean redshift of galaxies in the SDSSGC. For source galaxies, we
consider two models. The first model (hereafter LEV1) is for SDSS-like
surveys, which, to a certain extent, represents the current
state-of-the-art for large surveys.  The second model (hereafter LEV2)
is for future LSST-like surveys.  For LEV1 and LEV2 the source galaxy
redshift is taken to be $z_s=0.3$ and $z_s=1$, respectively.  For
simplicity, we do not consider detailed redshift distributions for
source and lens galaxies.  This may lead to inaccurate predictions for
lensing signals, especially for LEV1 with relatively low
$z_s$. However, for the purpose of comparing the detectibillity of
LEV1 and LEV2 surveys, our simplification should be adequate. On the
other hand, for future studies requiring high precision, the redshift
distributions of source and lens galaxies has to be properly 
accounted for.

\begin{figure*}
\includegraphics{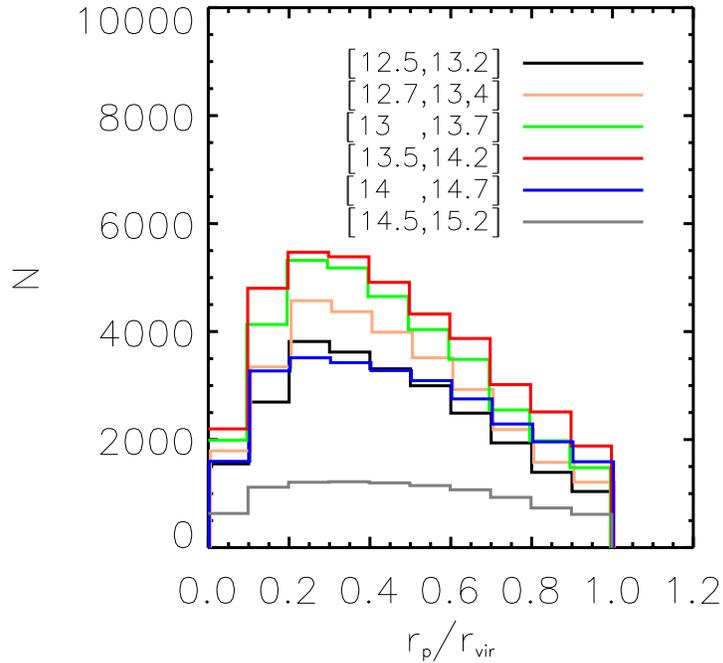}
\caption{The number of galaxies in SDSSGC as function
  of projected group-centric radius. Different line styles represent
  different host halo masses. The ranges of $\log(M/\ms)$ are
  marked in the figure.  }
\label{fig:rp}
\end{figure*}

With fixed $z_l$ and $z_s$, we have
\begin{equation}
  \sigma_{\Delta\Sigma}(R)=\sigma_{\gamma}(R)\times\Sigma_{crit}(z_l,z_s)\,.
  \label{dsig}
\end{equation}
Thus the lensing measurement noise only depends on $N_{\rm pair}$,
which in turn is determined by the number of lens galaxies and the
number density of source galaxies. We use the SDSSGC to estimate the
number of lens galaxies. Fig.\ref{fig:rp} shows the number of lens
galaxies in SDSSGC as a function of the projected group-centric radius
for different halo mass ranges.  Typically the SDSSGC provides
between 2000 and 6000 satellite galaxies (= lenses) per bin in host
halo mass and group-centric radius, for the binning adopted
here.  When halo mass decreases, the number of groups keeps 
increasing, but the number of satellites per host goes down.
At halo mass range of $10^{13} - 10^{14} \ms$ , one obtains largest number of
lens galaxies.

Future surveys such as LSST will not include spectroscopic data, making it
difficult to construct a reliable group catalog from the survey data
itself. This will have to await future deep and wide spectroscopic
surveys which will allow the construction of very large group
catalogs, out to high redshifts. This would allow the galaxy-galaxy
lensing based subhalo studies proposed here to be extended to higher
redshifts, using tomography. An alternative is to use
photometric redshifts for the construction of group catalogs. We
leave these avenues for future investigation. In what follows we
simply adopt the lens statistics from the SDSSGC for both LEV1 and
LEV2; i.e., even for LEV2, we still only consider groups constructed
from relatively shallow SDSS-like spectroscopic surveys.  Finally, for
the number density of source galaxies we adopt $1.6\hbox{ }{\rm
  arcmin^{-2}}$ for LEV1 \citep[as appropriate for SDSS,
  see][]{Mandelbaum2005} and $60\hbox{ }{\rm arcmin^{-2}}$ for LEV2.

\begin{figure*}
\begin{center}
\includegraphics[width=0.7\textwidth]{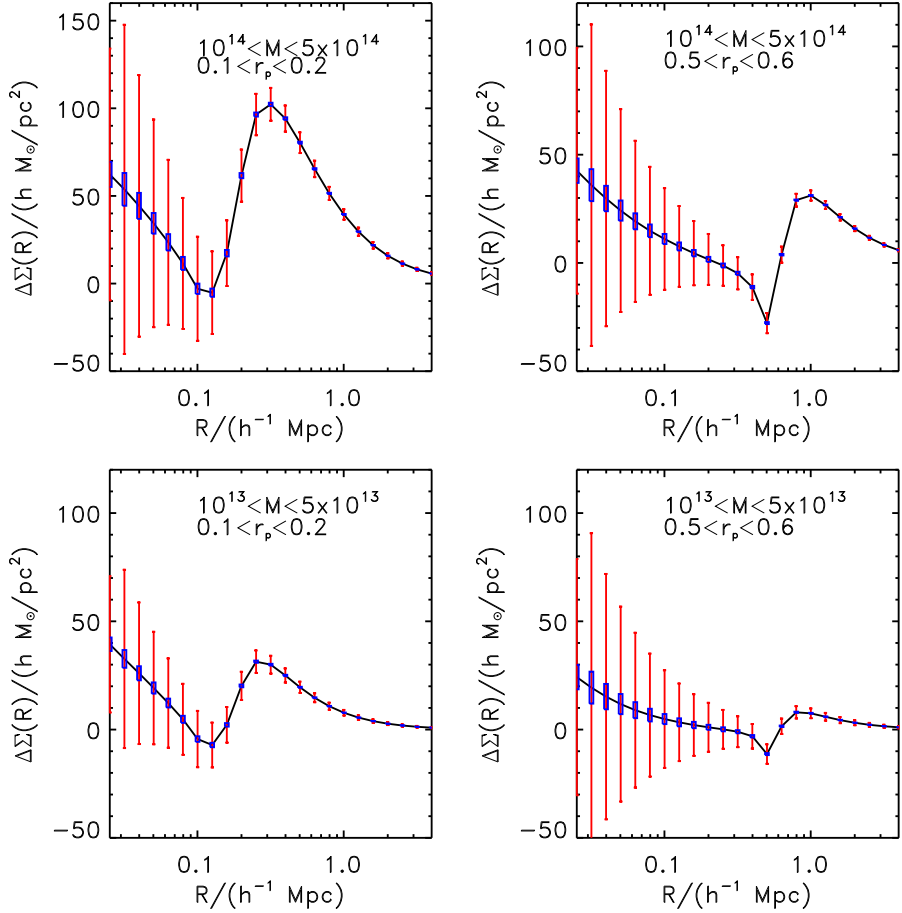}
\end{center}
\caption{The galaxy-galaxy lensing signal around
satellites in certain host halo mass bins and projected distance bins 
compared with the noise estimation. The red errorbars show the noise
estimation of LEV1 (SDSS like survey), while the blue rectangles show the 
LEV2 noise (LSST-like survey).  }
    \label{fig:sn}
\end{figure*}

In Fig.\ref{fig:sn}, we compare the lensing measurement noise for LEV1
and LEV2, respectively.  We calculate average lensing signals around
satellite galaxies with host halo mass in the range of
$[10^{14},5\times10^{14}]\ms$ and $[10^{13},5\times10^{13}]\ms$ and
with halo-centric distance in the range of $[0.1, 0.2]$ and $[0.5,
  0.6]$ $\mpch$.  We plot the expected noise from our two noise
models, LEV1 and LEV2, with red and blue error bars, respectively.  It
is seen that both the SDSS-like survey and the LSST-like survey can
detect the lensing signal from the host halo well.  For the inner
parts where the subhaloes dominate, the observational noise from
SDSS-like survey is much larger than the signal. On the other hand, a
LSST-like survey can detect the signal with high S/N. The S/N does not
drop for smaller groups, because the number of smaller groups is much
larger than that of massive ones and the mean subhalo mass does not
drop significantly in smaller groups (see Fig.\ref{fig:msdis}).

\section{Model inference with the Markov Chain Monte Carlo method}
\label{sec:mcmc}

\subsection{Markov Chain Monte Carlo fitting}

For a given set of observational data ${\bf \theta}$, the posterior
probability of the model parameters ${\bf \pi}$ can be derived from
the likelihood function, $\mathcal{L}({\bf \theta}|{\bf \pi})$, and
the prior probability, $P({\bf \pi})$, of these parameters. According
to Bayes' rule, we can write
\begin{equation}
  P({\bf \pi}|{\bf \theta})={ \mathcal{L}({\bf \theta}|{\bf \pi})
  P({\bf \pi})\over P({\bf \theta)}}\,,
  \label{eq:bayesian}
\end{equation}
where the normalization, $P({\bf \theta)}$, is called the evidence.
The prior probability $P({\bf \pi})$ describes our known knowledge
about the model parameters.  In our fiducial computation, we simply
adopt a flat prior over a range in the parameter space. As a test of
the robustness of our inferences, we also use another set of priors.
We write the likelihood function $\mathcal{L}$ as
\begin{equation}
  \ln \mathcal{L}=\sum_{i} \left( \frac{\Delta\Sigma(R_i|{\bf
  \pi})-{\Delta\hat{\Sigma}(R_i)}}{\sigma_{\Delta\Sigma}}\right)^2 \,,
  \label{eq:L}
\end{equation}
where ${\Delta\hat{\Sigma}({R}_i)}$ is the observed excess surface
density at radius bin $R_i$, $\Delta\Sigma(R_i|{\bf \pi})$ is the
theoretical prediction with model parameters ${\bf \pi}$, 
and $\sigma_{\Delta\Sigma}$ is the error estimate given by 
equation (\ref{dsig}). In this
paper, we use MCMC to explore the posterior distribution $P({\bf
  \pi}|{\bf \theta})$.  The key component of the MCMC method is a
guided random walk in the parameter space. The likelihood function at
a certain volume of the parameter space is then proportional to the
number density of points in that volume. The MCMC sampler used here is
that provided in the CosmoMC package \citep{lewis2002}, which adopts
the Metropolis-Hastings algorithm \citep{Metropolis49, Metropolis53,
  Hastings70} by default.  A detailed review of this method can be
found in \cite{Chib1995}.  For each fitting process, we generate three
MCMC chains starting from different positions in the parameter
space. We use the runtime convergence criteria in CosmoMC, which
computes the standard Gelman and Rubin $R$-statistic diagnostics to
monitor the convergence.  We declare convergence when $R<1.1$. Only
the second half of the chain (which is well converged) 
values are used to sample the posterior probability.

\subsection{Model inference}\label{sec:result}

\begin{table}
\caption{The mean values of input model properties, 
and the ranges of the parameter values adopted 
as priors in the MCMC. $M$ and $M_{\rm sub}$ are in units 
of $\ms$; $r_{\rm p}$ and $r_{\rm s,sub}$ are in units of 
$\mpch$; $\rho_{\rm 0,sub}$ is 
in units of $10^{16}h^2M_{\odot} {\rm Mpc}^{-3}$.}

\begin{tabular}{|c|c|c|c|c|c|c|}
 \hline & $\log{M}$  & c & $r_{\rm p}$ & $\log{M_{\rm sub}}$ & $\rho_{\rm
   0,sub}$ & $r_{\rm s,sub}$ \\
  \hline
  \hline
 mean input& 14.247 & 6.89 & 0.55 & 11.67& 0.99 & 0.0155 \\ 
 \hline
 high bound& 14.5  & 10  & 0.6   & 12   & 10   & 0.2 \\
   \hline
 low  bound& 13.5  &  3  & 0.5   & 9   &  0.1   & 0.001\\
 \hline
\end{tabular}
\label{tab:para}
\end{table}

Here we investigate to what extent the observations of satellite
galaxy-galaxy lensing from LEV1 and LEV2 surveys can constrain the
satellite and host halo properties described by a set of parameters,
including $(M, c, r_p)$ for the host halo and $(M_{\rm sub}, \rho_{\rm
  0,sub}, r_{\rm s,sub})$ for the subhalo.  We construct the
`observed' galaxy-galaxy lensing signals for satellites selected from
SDSSGC following the descriptions in \S4. We consider separately two
subsamples of satellites according to their host halo mass, one in the
range of $[10^{14}, 5\times10^{14}]\ms$ (SAMPLE1), and the other for
$[10^{13},5\times10^{13}]\ms$ (SAMPLE2). The projected halo-centric
distances of satellites are chosen to be in the range of $[0.5, 0.6]
\mpch$ for both subsamples. We stack the galaxy-galaxy lensing signals
for the satellite galaxies in each subsample thus constructing a set
of `observed' signals. We then employ MCMC fitting to these mock data
in order to derive constraints on the mean values for the six
parameters, three for host haloes $(M, c, r_p)$ and three for subhaloes
$(M_{\rm sub}, \rho_{\rm 0,sub}, r_{\rm s,sub})$.  The selection of
the bin size in the projected halo-centric distance of satellites is a
compromise between statistical errors and systematic bias. If we used
the bin size of $0.2 \mpch$ instead of $0.1 \mpch$, for example, the
source galaxy number would increase by about a factor of two and thus
the statistical errors are decreased.  However, stacking
satellites over a larger range in $r_{\rm p}$ leads to larger bias
in the derived mean host halo mass $M$ and concentration parameter $c$
from MCMC fitting to the stacked satellite galaxy-galaxy lensing
signals.

To derive the parameters we are interested in, we fit the mock lensing
data with the model,
\begin{equation}
\Sigma(R)=\Sigma_{\rm host}(R|M,c,r_p) + \Sigma_{\rm
  sub}(R|M_{\rm sub},\rho_{\rm 0,sub},r_{\rm s,sub})\,,
\end{equation}
where
\begin{equation}
\Sigma_{\rm host}(R|M,c,r_p)=\frac{1}{2\pi}\int_0^{2\pi}
\Sigma(\sqrt{r_p^2+R^2+2r_pR\cos{\theta}}) d\theta,
\end{equation}
where $\Sigma(R)$ is the projected density profile of the NFW host
halo with mass $M$ and concentration parameter $c$. Note that $R$ is
the distance to the satellite galaxies around which we detect
galaxy-galaxy lensing signals. The second term $\Sigma_{\rm sub}(R)$
is the projected density of the subhalo with profile:
\begin{equation}
\rho_{\rm sub}= \left\{
\begin{array}{rl} \frac{\rho_{\rm 0, sub}}{(1+r/r_{\rm s,sub})^2(r/r_{\rm
    s,sub})} &\text{if } r < r_t \\
    0 & \text{if } r > r_t
\end{array}\right.\,.
\end{equation}
where the truncation radius, $r_t$, is computed using
Eq.~(\ref{eq:rt}).

We first analyse SAMPLE1 with relatively massive host haloes.
Table~\ref{tab:para} shows the mean values of the six parameters of
the input sample, and the boundary of the flat priors we adopt in our
MCMC fitting.

\begin{figure*}
\includegraphics[width=\textwidth]{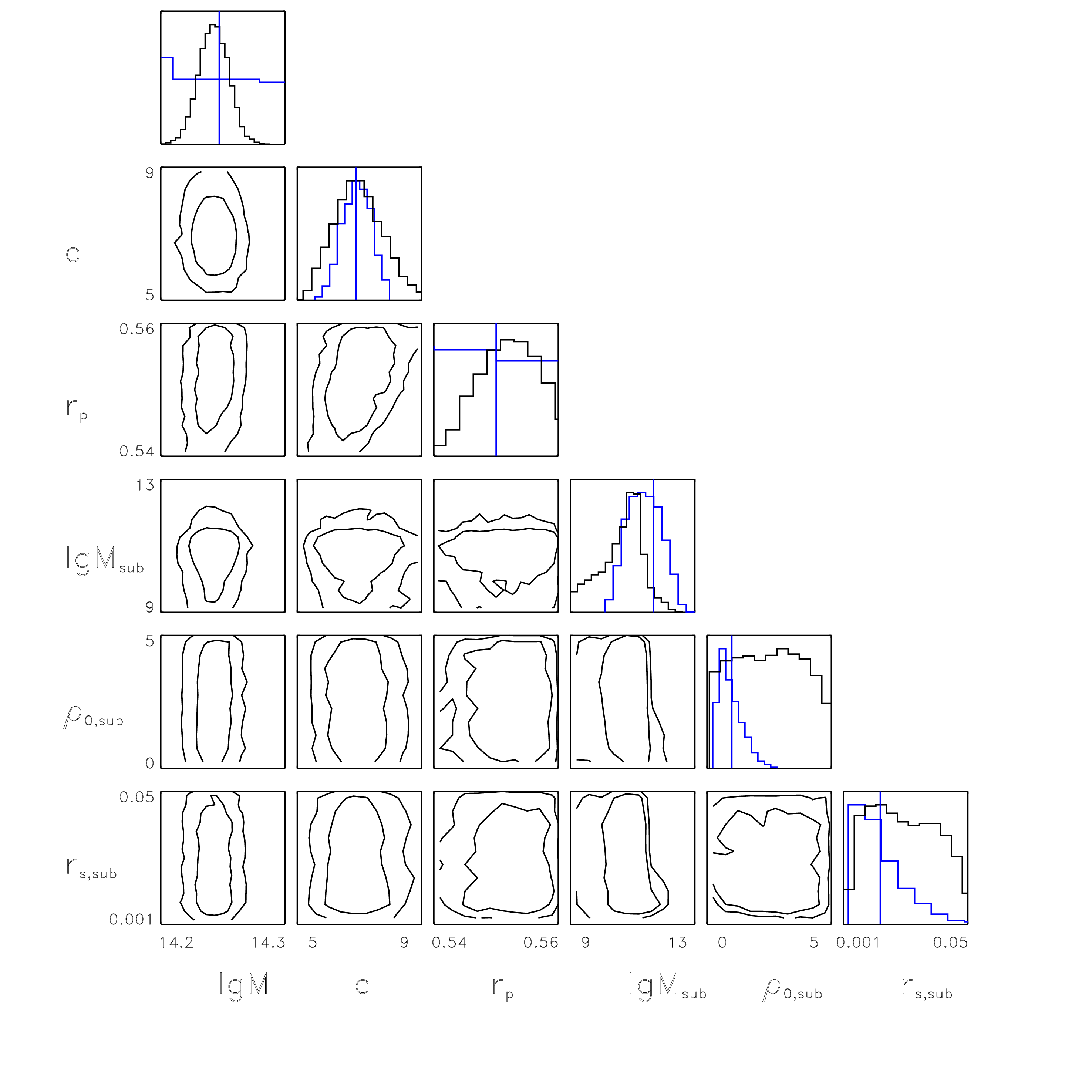}
\caption{The marginalized posterior probability distribution for the
  model parameters in LEV1 case. The contours show $68\%$ and $95\%$
  confidence levels. The last panel of each row shows the 1-d
  marginalized probability distribution, together with the average
  value of the input satellites (vertical solid lines).  The blue
  histograms show the corresponding distributions obtained directly 
from the input SAMPLE1.}
\label{fig:MCMC_SDSS}
\end{figure*}
\begin{figure*}
\includegraphics[width=\textwidth]{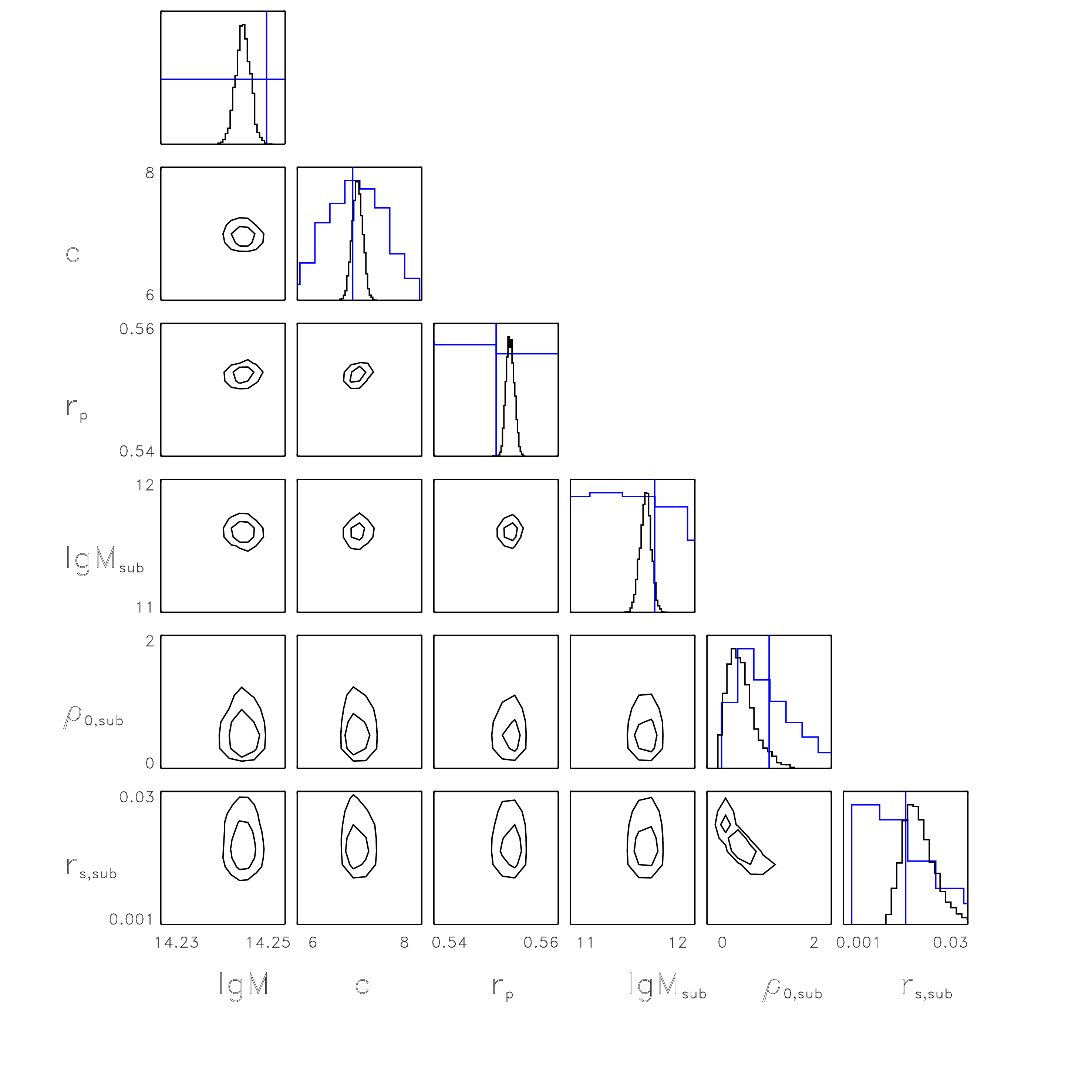}
\caption{The same as Fig.\ref{fig:MCMC_SDSS} but for results using
  LEV2 noise model.}
\label{fig:MCMC_lsst}
\end{figure*}

We separately analyse the constraints expected from LEV1 surveys and
LEV2 surveys.  Fig.\ref{fig:MCMC_SDSS} shows the marginalized
posterior probability distribution for the model parameters for
LEV1. The contours indicate the $68\%$ and $95\%$ confidence
levels. The last panel in each row shows the marginalized probability
distribution for the corresponding parameter, in which the probability
distribution (blue histogram) and the average value (vertical line)
from the input sample are also shown for comparison. As is evident,
even for LEV1 the host halo mass $M$ and concentration parameter $c$
can already be constrained reasonably well. However, the constraints
on the subhalo mass are extremely weak, with the $68\%$ confidence
range covering two orders of magnitude. For the subhalo density
profile, no meaningful constraints can be obtained from LEV1-type
surveys.
  
The results obtained using the LEV2 noise model are shown in
Fig.\ref{fig:MCMC_lsst}. Clearly, the factor 50 increase in the number
density of source images causes a dramatic improvement in the
constraints on the model parameters compared to those in
Fig.\ref{fig:MCMC_SDSS}. The host halo mass and the concentration
parameter can be constrained with high precision.  The subhalo mass
can also be tightly constrained with a $1\sigma$ confidence range of
about $0.2$ dex.  For the amplitude and scale radius of the subhalo
density profile, reasonable constraints can be achieved.  Note,
though, that there is a strong degeneracy between $\rho_{\rm 0,sub}$
and $r_{\rm s,sub}$, as seen from their joint constraints in the fifth
panel of the bottom row.

\begin{figure*}
\includegraphics[width=\textwidth]{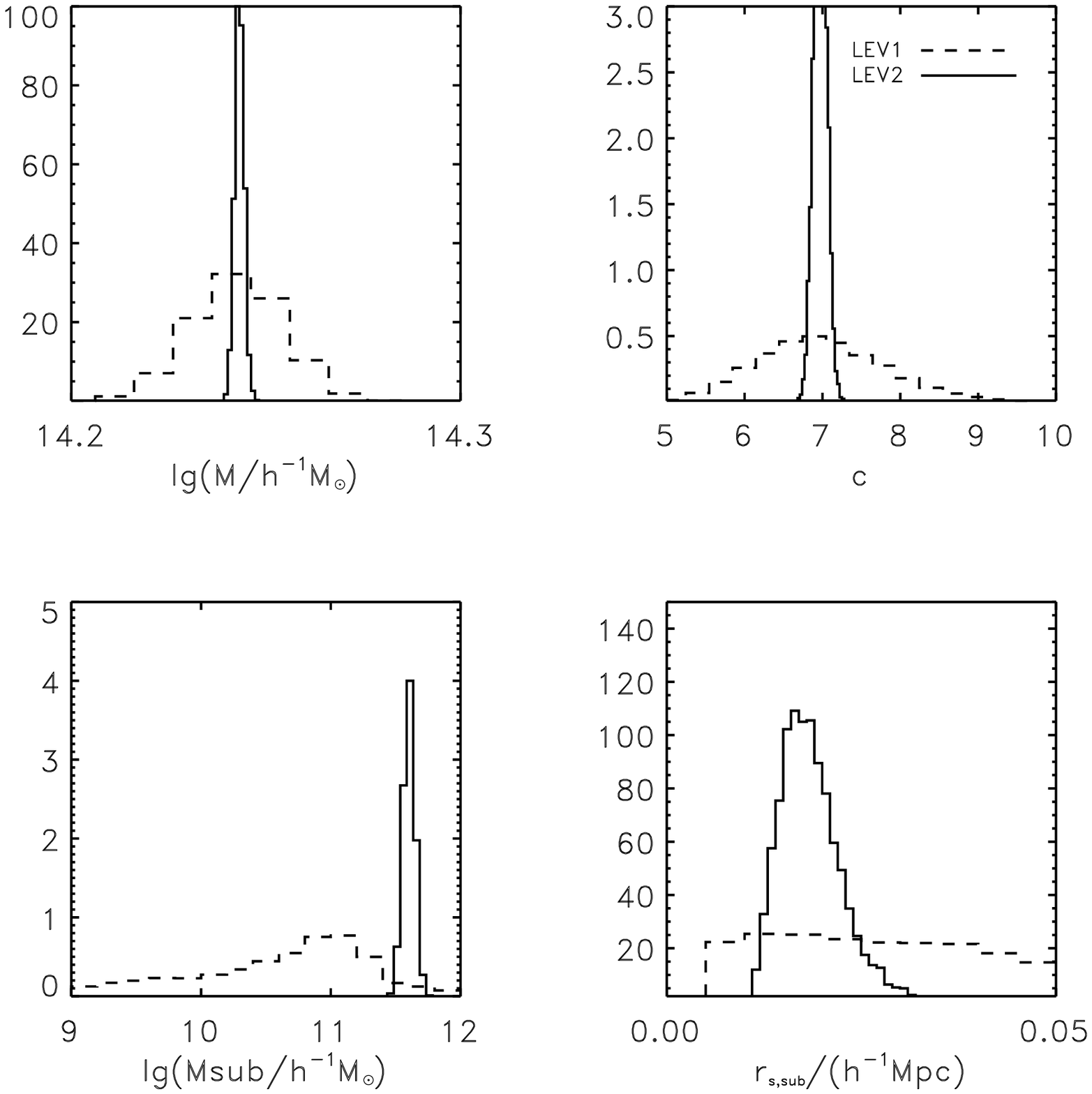}
\caption{A close comparison between the posterior distributions 
shown in Figs.\ref{fig:MCMC_SDSS} and \ref{fig:MCMC_lsst}. The solid 
histograms are the marginalized distribution of $M$, $c$, 
$M_{\rm sub}$ and $r_{\rm s, sub}$ for LEV2, while the dashed histograms 
for LEV1.}
\label{fig:compare}
\end{figure*}
\begin{figure*}
\includegraphics[width=\textwidth]{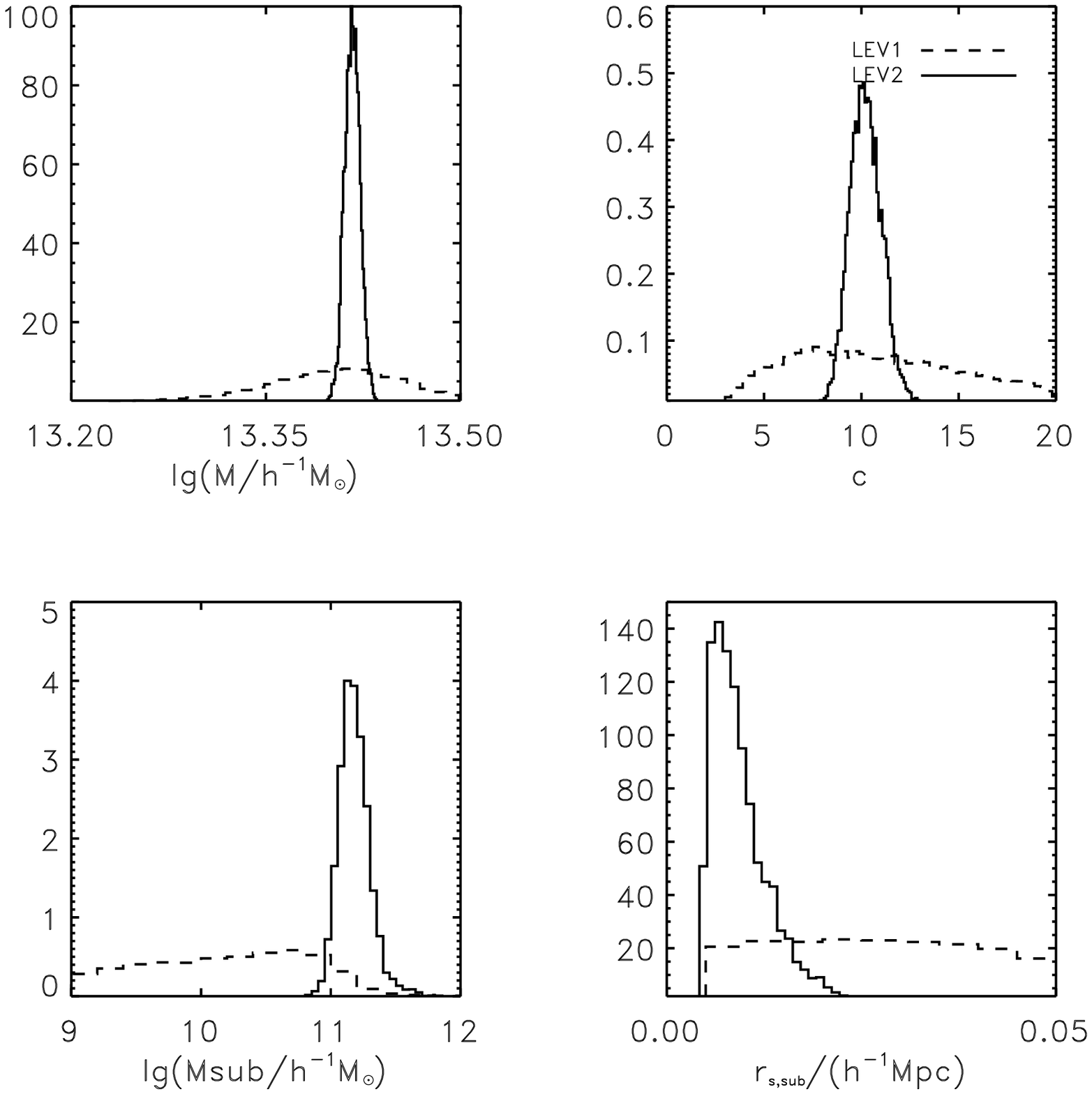}
\caption{The same as Fig.\ref{fig:compare}, except that here results 
are shown for satellites in host haloes with masses in the 
range $[10^{13},5\times 10^{13}]\ms$ and with projected halo-centric 
distance in the range $[0.5,0.6]\mpch$.}
\label{fig:compare_mbin1}
\end{figure*}

To see more clearly the improvements from LEV1 to LEV2, in
Fig.\ref{fig:compare} we directly compare the marginalized probability
distributions for $M$, $c$, $M_{\rm sub}$ and $r_{\rm s, sub}$
obtained using the LEV2 (solid) and LEV1 (dashed) noise models.  This
demonstrates the remarkable potential of the next generation of
LSST-like surveys for subhalo studies using satellite galaxy-galaxy
lensing.

We also perform the MCMC fitting for SAMPLE2 with group sized host
haloes with mass in the range of $[10^{13},5\times 10^{13}]$ $\ms$.
The results are shown in Fig.\ref{fig:compare_mbin1}.  It is seen that
for group sized host haloes and the satellites therein, we can still
get good constraints with LEV2 surveys.

\subsection{The impact of prior choice}
\label{sec:prior}

\begin{figure*}
\begin{center}
\includegraphics[width=0.9\textwidth]{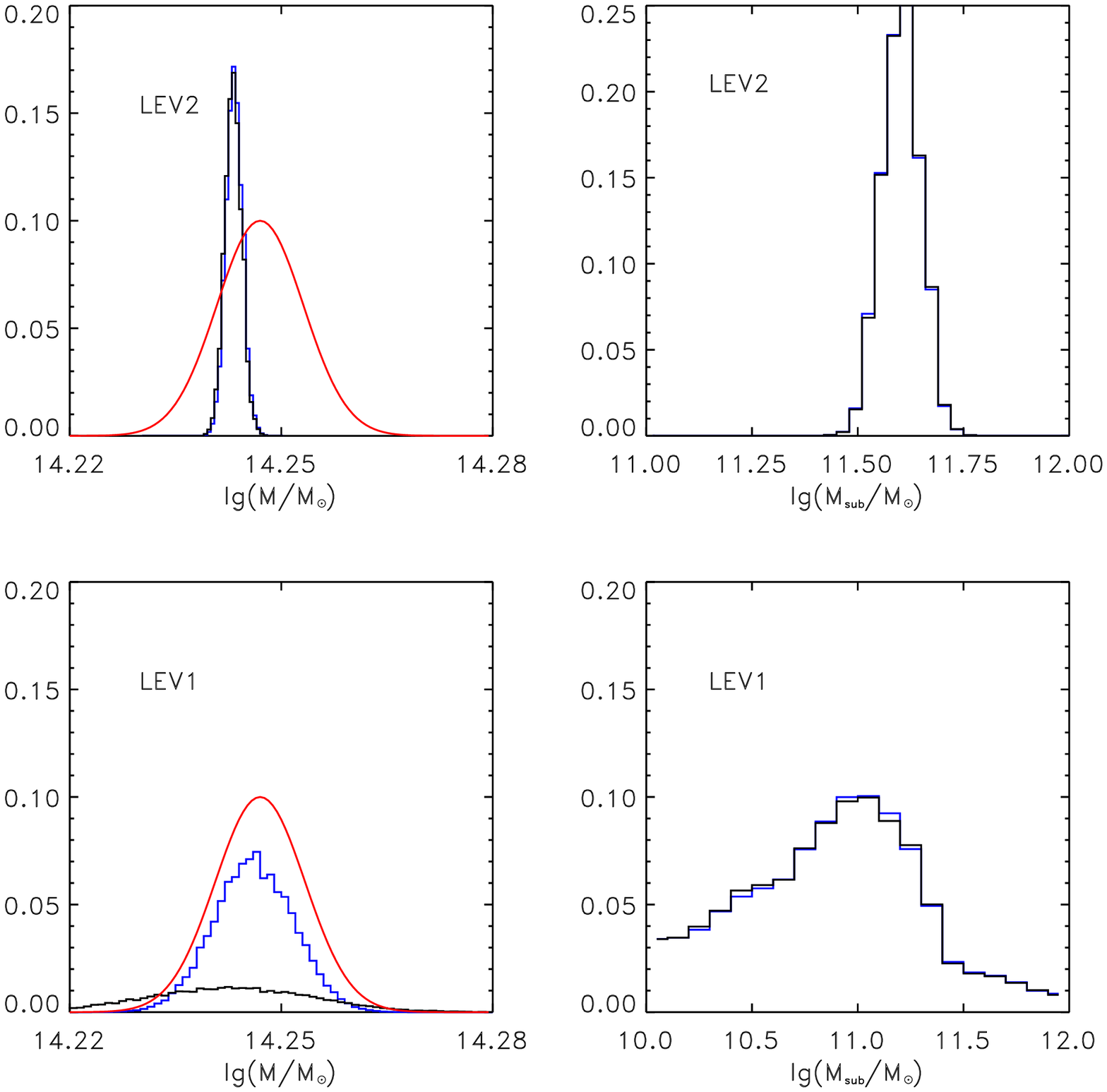}
\end{center}
\caption{The figure shows the effect of priors of host halo mass on
parameter inferences.The black (blue) histogram represents the
marginalized posterior distribution of $M$ (left column) and 
the subhalo mass (right column) with flat prior (lognormal prior). The upper
 panels show the results for LEV2 while the lower panels show the
results for LEV1. In the left panels, the red solid lines show
the prior distribution of halo masses. We scale the amplitude of 
the prior distribution so that it can be shown in the same panel 
together with the posterior distribution. }
\label{fig:prior1}
\end{figure*}

The results above are obtained with flat priors for all
parameters. Here we test the impact of prior choice on our inferences
of model parameters. Since the groups used here are selected from
SDSSGC group catalog, each group has already been assigned an
estimated mass. Thus we do have some idea about the mass distribution
of the selected groups that may be used as priors in the MCMC fitting.
As an approximation, we model the mass distribution with a log-normal
function,
\begin{equation}
P( \log{M} )= \frac{1}{\sqrt{2\pi}\sigma} 
\exp\left({-\frac{ (\log{M} - \log{\bar{M})^2 } }{2\sigma^2} }  \right) \,,
\end{equation}
where $\bar{M}$ is the mean mass of the selected groups, and
$\sigma=(\sigma_0 + \sigma_1)/\sqrt{N}$, with $\sigma_0=0.3$ the mass
assignment uncertainty in the SDSSGC, $\sigma_1$ the standard
deviation of the group mass in logarithmic space, and $N$ the number
of selected groups. In Fig.\ref{fig:prior1}, we show the constraints
using this distribution as the prior for the host halo mass. The left
panels show the marginalized posterior distribution of $M$, and the
right panels are for the subhalo mass $M_{\rm sub}$.  The upper and
lower panels are for LEV2 and LEV1, respectively.  The black and blue
histograms are the results with flat priors for all the parameters,
and the lognormal prior for $M$ and flat priors for the other
parameters, respectively.  The red lines in the left panels illustrate
the lognormal prior distribution for $M$. For LEV2, because the
constraints are already tight, adding the lognormal prior on $M$ does
not change the constraints significantly, although it does decrease
the bias in $M$ by a small (barely significant) amount.  In the case
of the LEV1 noise model, the posterior distribution for $M$ is
essentially identical to its prior distribution, indicating that the
LEV1 lensing data does not improve the constraints on host halo mass
beyond our prior knowledge.  For the subhalo mass, the posteriors
based on both LEV1 and LEV2 are not affected by the prior on $M$. This
is expected from Figs.\ref{fig:MCMC_SDSS} and \ref{fig:MCMC_lsst},
which show that the degeneracies between host halo mass and subhalo
parameters are very weak.
                     
\section{Bias and contaminations}
\label{sec:discussion}

\subsection{Stacking bias}

From Figs.\ref{fig:MCMC_SDSS} and \ref{fig:MCMC_lsst}, it can be seen
that the peak of the posterior distribution for some parameters
deviates from the input mean value. These biases are from fitting the
model to galaxy-galaxy lensing data obtained by stacking a large
sample of satellite galaxies..  The properties of the satellite
galaxies and their host haloes in the sample are not identical but
spread over certain ranges. The parameters obtained from the fitting
to the stacked signals then correspond to the results of certain
averages over such a sample.  Depending on the quantities, the
averages obtained may be biased relative to the means of the input
values. In our analyses, although the properties of lens galaxies,
such as their host halo mass and the halo-centric locations, are
selected to be similar, they still cover finite ranges with some
distributions. In particular the subhalo mass of the lens galaxies
covers a very broad range. Bias arises simply because the lensing
signal depends on these parameters in a non-linear fashion.  For
subhalo mass, the difference between the peak of the posterior
distribution and the input mean value is about 0.1 dex, slightly
larger than the $2\sigma$ width of the posterior distribution.

\begin{figure*}
 \begin{center}
 \includegraphics[width=\textwidth]{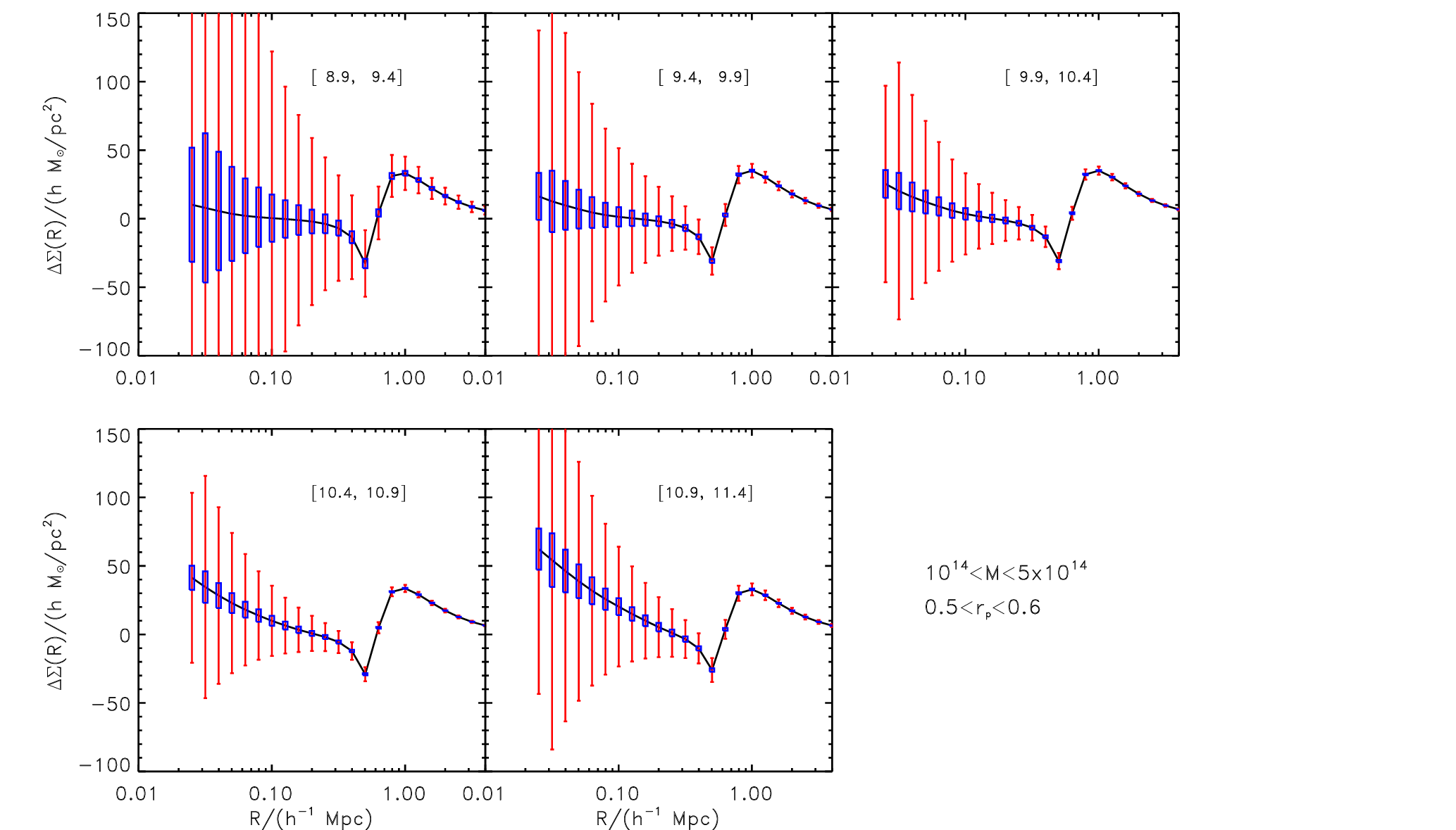}
 \end{center}
 \caption{The lensing signal around satellite galaxies in different 
 stellar mass range. All satellite galaxies are selected from groups of
 $[10^{14},5\times10^{14}]\ms$, and with projected halo centric distance 
 of $[0.5, 0.6]\mpch$. Each panel shows lensing signal of satellites
 in certain stellar mass bin. We mark the $\log(M_*/\ms)$ range for each 
 subsample in each penel. The red errorbars show the noise  estimation 
 of LEV1 (SDSS like survey), while the blue rectangles show the  LEV2 
 noise (LSST like survey). } 
 \label{fig:snbin}
\end{figure*}

\begin{figure*}
 \begin{center}
 \includegraphics[width=\textwidth]{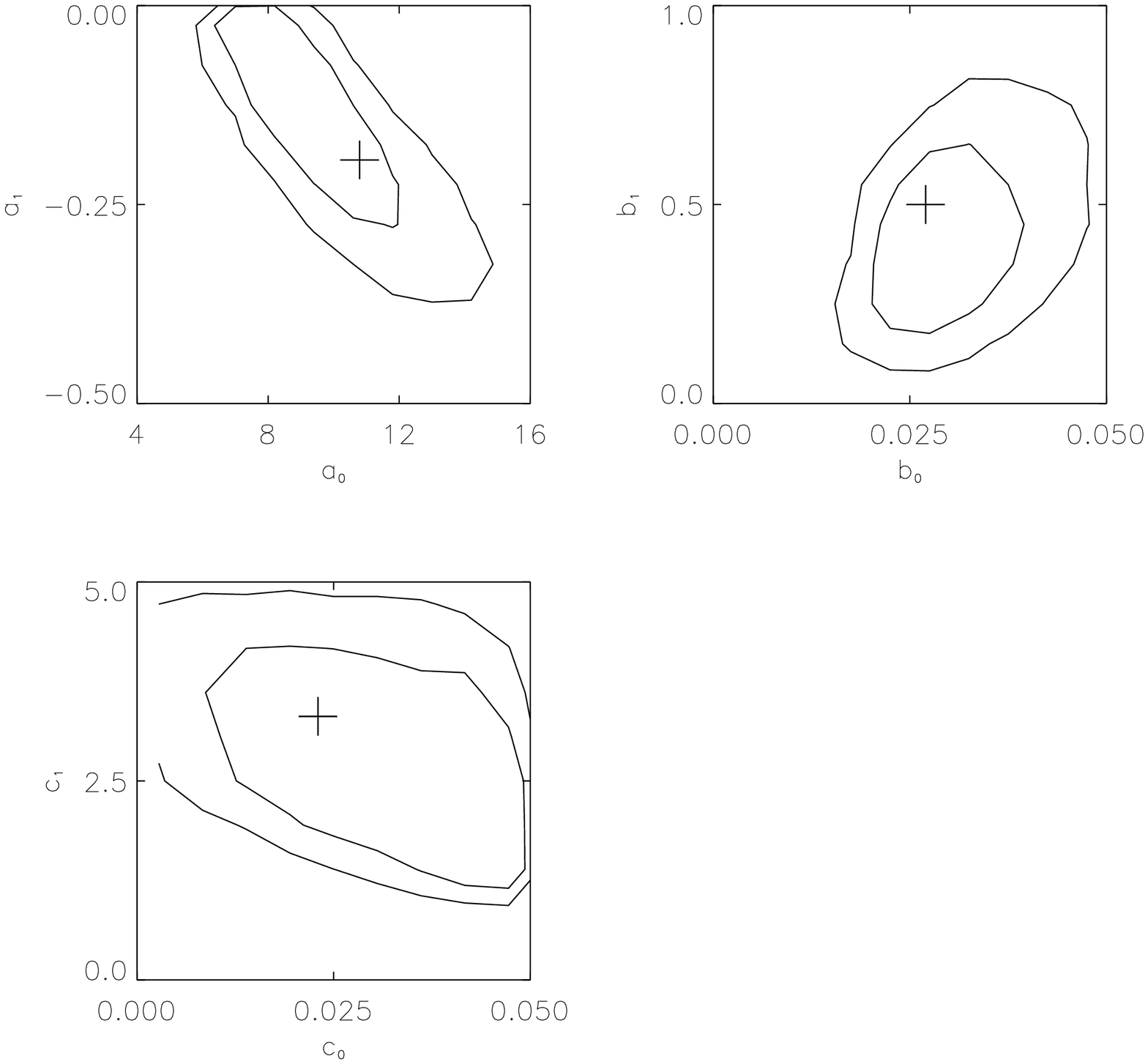}
 \end{center}
 \caption{Fitting results for lensing signals in Fig.\ref{fig:snbin} with LEV2 noise. 
 Contours show 68\% and  95\% confidence range of the marginalized 
 posterior probability distribution. The crosses mark 
 the true input values.}
 \label{fig:9p_2d}
\end{figure*}

One way to suppress the bias is to use narrower mass bins.  In our
model, the subhalo mass of a satellite depends mainly on its stellar
mass although the host halo mass and halo-centric radius also
influence somewhat through tidal interactions.  To see the effects of
different binning, we split satellites in our fiducial satellite
sample into five finer stellar mass bins. The corresponding lensing
signal of each of the mass bins is shown in Fig.\ref{fig:snbin},
respectively. We then can analyze the host halo and subhalo properties
with MCMC by fitting to the lensing signals from the five subsamples
jointly. In principle, we should consider five sets of parameters
$(M_{\rm sub}, r_{\rm s,sub}, \rho_{\rm 0,sub})$ each for a single
bin. We then need to deal with $15$ parameters for subhaloes, plus the
ones for host haloes.  The task would be difficult with the statistics
of the expected data. On the the hand, one naturally expects certain
relations between different quantities, which can be parameterized
with a much smaller number of free parameters. By MCMC fitting to the
`observed' data, we can extract the constraints on these
parameters. This approach allows us to control the number of free
parameters and at the same time to model the lensing signals better
than that from a single broad bin of subhalo mass. As a test, we
assume that $M_{\rm sub}/M_*$ depends on the stellar mass $M_*$
according to a power-law,
\begin{equation}
\label{eq:ML}
\frac{M_{\rm sub}}{M_*} =a_0 \left ( \frac{M_{*}}{{10^{10}\ms}}\right)^{a_1}\,.
\end{equation}
Furthermore, we assume that $r_{\rm s,sub}$ and $r_{\rm t}$ 
depend on subhalo mass through the following parameterizations
\begin{equation}
\label{eq:rs_msub}
r_{\rm s,sub}=b_0 \left ( \frac{M_{\rm sub}}{{10^{12}\ms}}\right)^{b_1}\,,
\end{equation}
and
\begin{equation}
\label{eq:rcut_msub}
r_{\rm t}=c_0+c_1\,r_{\rm s,sub}\,,
\end{equation}
where $a_0$, $a_1$, $b_0$, $b_1$, $c_0$ and $c_1$ are all free parameters.
We then fit the mock lensing data from all $5$ stellar mass bins
simultaneously to derive constraints on these parameters.
Fig.\ref{fig:9p_2d} shows the 68\% and 95\% confidence range of the
marginalized posterior probability distributions for these subhalo
parameters. For comparison, the `true input' values, obtained by
directly fitting relations~(\ref{eq:ML}), (\ref{eq:rs_msub}), and~
(\ref{eq:rcut_msub}) to the input subhaloes, are marked with the plus
symbol in each panel. We can see that the 68\% posterior contours
enclose the input values, indicating that the bias due to binning can
be effectively reduced by dividing lens galaxies into fine bins.

\subsection{Contamination from fake group members}

So far we have not considered possible contaminations in the group
catalog and assumed that all galaxies assigned to a group are true
members. In reality, however, some galaxies that are identified as
satellites may actually be central galaxies of other (low-mass) haloes
along the line of sight. In what follows, we refer to such galaxies as
interlopers. For galaxies of the same luminosity, central galaxies
produce much stronger lensing signals than that of satellites. Hence,
even an interloper fraction of 10\% can introduce significant errors
in the inferred subhalo parameters.
\begin{figure}
 \begin{center}
 \includegraphics[width=0.5\textwidth]{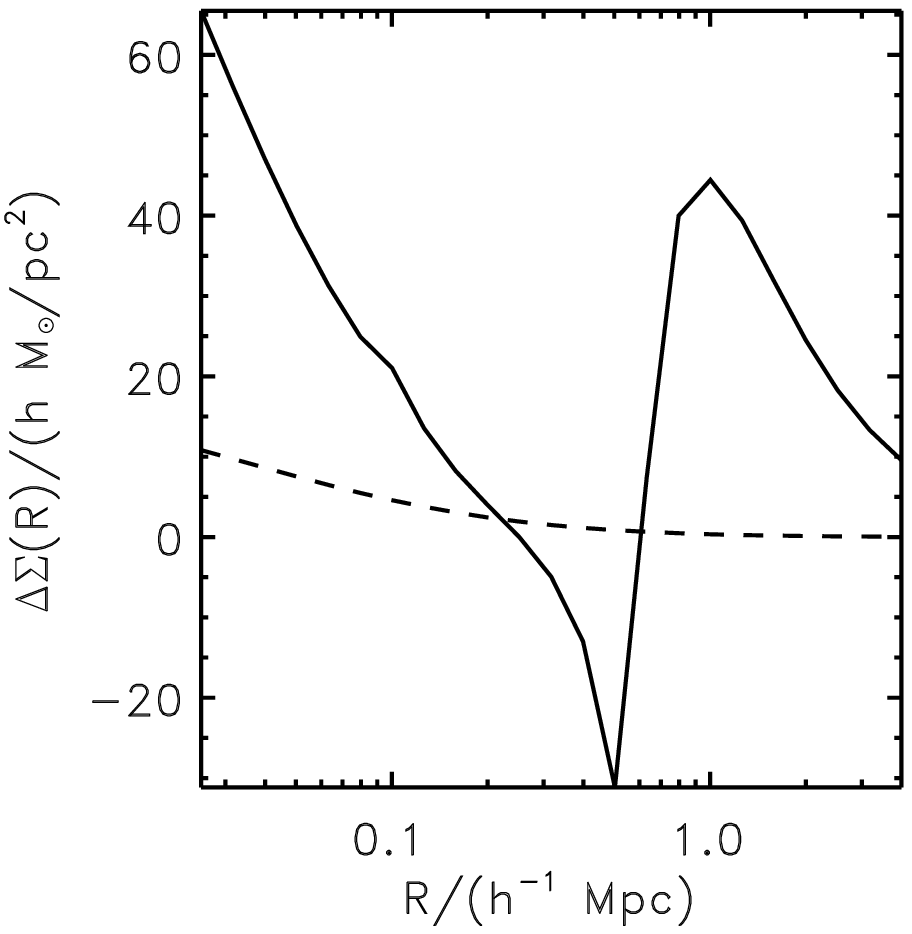}
 \end{center}
 \caption{The lensing signal of  galaxies in the mock SDSS 
  group catalog. The galaxies are selected from haloes of   
  $[10^{14},2\times10^{14}]\ms$ with halo-centric radius of $[0.5,0.6]\mpch$. 
  The signal from true satellites is represented by solid line and 
  that from fake members by dashed line.} 
 \label{fig:con}
\end{figure}
\begin{figure}
 \includegraphics[width=0.5\textwidth]{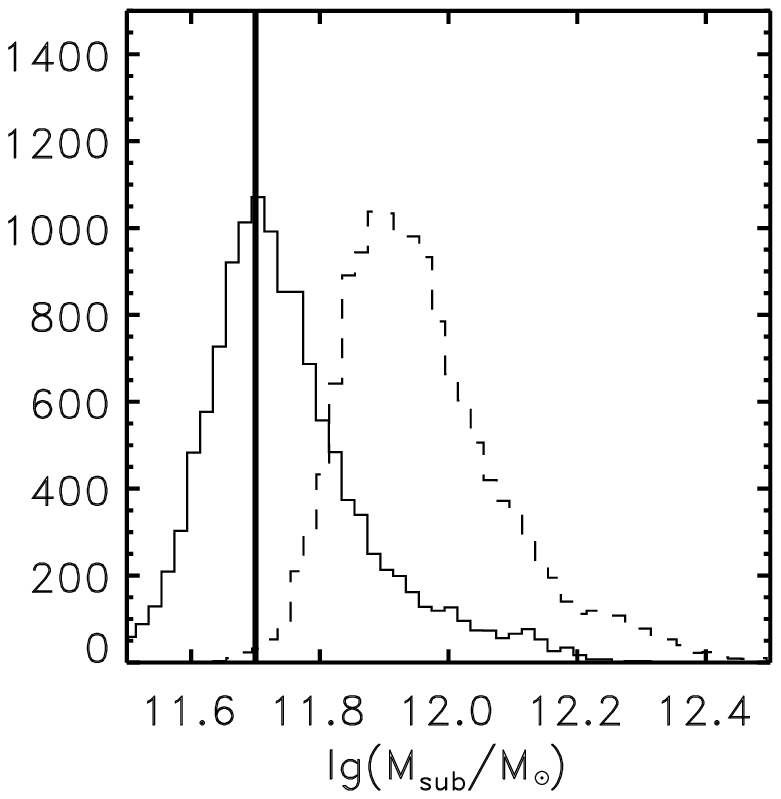}
 \caption{The 1-d constraint on the subhalo mass. The dashed and solid
   lines are for the results derived from fitting to the `mixed' and
   `true' signals of satellites, respectively.  The vertical line
   indicates the input value of the subhalo mass.}
\label{fig:mock}
\end{figure}

To estimate the impact of interlopers, we make use of the SDSS mock
group catalog provided by \citet{Yang2007}. This mock catalog is
constructed by running the halo-based group finder of \citet{Yang2005}
on a mock galaxy redshift survey (MGRS) built by populating dark
matter haloes with galaxies according to the conditional luminosity
function\citep[CLF;][]{vandenBosch2007}.  The CLF, which describes the
luminosity distribution for galaxies in haloes of a given mass, is
constrained using the clustering and abundances of galaxies in the
SDSS.  As the result, the luminosity function and clustering
properties of the MGRS accurately matches those of the SDSS.  .  The
MGRS also takes into account of real observational conditions by
mimicking the sky coverage and completeness trend of the SDSS survey
\citep[see][for details]{Yang2007}.  For such a mock group catalog,
we not only know the group to which a galaxy is assigned, but also the
dark matter halo to which the galaxy truly belongs. Thus, it is
particular suitable to examine the potential bias that arises due to
interlopers.
  
From the SDSS mock group catalog, we select satellites in groups with
assigned host mass from the group finder in the range of $[10^{14} ,
  2\times10^{14}]\ms$ and with the projected halo-centric distance of
$[0.5 , 0.6] \mpch$ . We use the model described in Section
\ref{sec:model} to generate mock galaxy-galaxy lensing data, with LEV2 noise.
 To isolate the errors due to interlopers, we fix the subhalo mass to be
$10^{11.7}\ms$ in calculating the fiducial data, which is similar to
the mean subhalo mass of SDSSGC satellites used in previous
sections. We find that about $10\%$ of the selected satellite galaxies
are actually centrals of other host haloes (i.e., are interlopers). We
then assign each of them a NFW profile according to their host halo
mass. The resulting lensing signals are shown in Fig.\ref{fig:con},
where the solid line is the excess surface density profile of the true
satellite galaxies, and the dashed line shows the signal due to the
interlopers. As can be seen, the interlopers contribute about 15\% of
the total signal in the inner part, which in turn can lead to large
bias in the model fitting.  Fig.\ref{fig:mock} shows the bias in
subhalo mass introduced by these interlopers. We fit two sets of
lensing signals separately. The first set contains contributions from
both true satellites and interlopers. This is referred to as the
`mixed' signal. The second set contains only true satellite galaxies,
and is referred to as the `true' signal.  The solid line is the
constraint from the `true' signals and the dashed line is from the
`mixed' signals. The vertical line indicates the input subhalo mass.
We can see that the subhalo mass inferred from the `mixed' signal is
biased high by $\sim 50\%$ compared to that inferred from the `true'
signal. The latter result is very consistent with the input value.

\begin{figure}
 \begin{center}
 \includegraphics[width=0.5\textwidth]{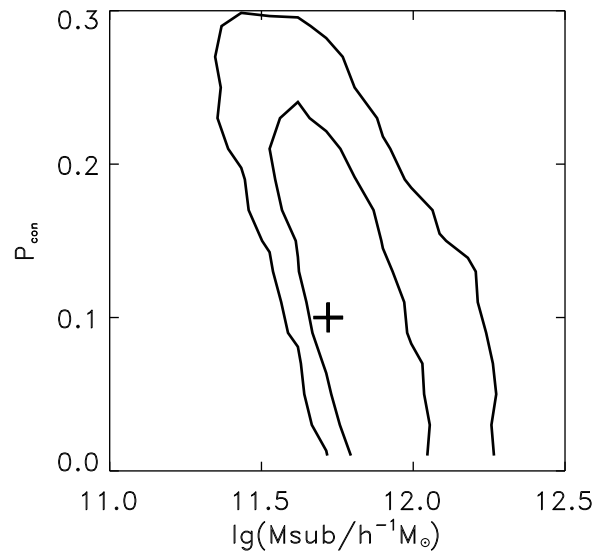}
 \end{center}
 \caption{Fitting to the mixed lensing signal of the mock catalog.  The
   contour shows 2d posterior distribution of $M_{\rm sub}$ and
   $P_{\rm con}$. The cross represents the input value. }
 \label{fig:9p_con}
 \end{figure}

The above analyses shows that it is important to carefully consider
the impact of interlopers in the group catalog. Unfortunately, it is
virtually impossible to completely eliminate interlopers.  In fact,
the halo based group finder of \citet{Yang2005} has been optimized to
minimize the impact of interlopers, among some other
constraints. Rather than trying to reduce (or eliminate) interlopers,
one may also try to account for them in the modeling. Using empirical
relations between the luminosity/stellar mass of central galaxies and
their host halo mass \citep[see e.g.][]{Yang2007} it is fairly
straightforward to fit for subhalo mass and interloper fraction
simultaneously. As a test, we assume that each galaxy in the selected
sample has the same possibility $P_{\rm con}$ to be a central galaxy.
We use the relation between the central galaxy luminosity and its host
halo mass from our MGRS to assign a halo mass to a central galaxy.
The total signal is then modeled as,
\begin{equation}
\Delta\Sigma(R)= \Delta\Sigma_{\rm sat}(R) (1-P_{\rm con} )+ 
  \Delta\Sigma_{\rm cen}(R) P_{\rm con}\,,
\end{equation} 
where $\Delta\Sigma_{\rm sat}$ is the lensing signal calculated by
assuming no contamination, and $\Delta\Sigma_{\rm cen}$ is the lensing
signal calculated by assuming that all satellites selected are
actually central galaxies of other haloes. The fitting result is shown
in Fig.\ref{fig:9p_con}. Although the posterior distribution shows
degeneracy between $P_{\rm con}$ and $M_{\rm sub}$, the bias due to
contamination is now suppressed.  If the MGRS is sufficiently
realistic, we should in principle be able to obtain some estimates for
the interloper fractions, which can then be used as a prior in the
MCMC fitting.  This should allow for tight and unbiased constraints on
the subhalo mass.

\subsection{Uncertainties due to the assumed halo density profile}

Our model presented above assumes that dark halo profiles are given by 
the NFW form. However, more recent investigations have shown that 
the  Einasto profile \citep{Einasto1965} represents CDM halos better
in the inner part \citep[e.g.][]{Navarro2004}. 
The Einasto profile can be written as
\begin{equation}
\rho(r)=\rho_{-2}\exp\left(-\frac{2}{\alpha}\left[\left(\frac{r}{r_{-2}}\right)^\alpha-1\right]\right)\,,
\end{equation}
where $\rho_{-2}$ and $r_{-2}$ are the density and radius where the local density
slope is $-2$. The parameter $\alpha$, usually called the shape
parameter, describes the change of the density slope with radius. 
With the state-of-art $N$-body simulations, it was found  that fixing
the shape parameter $\alpha \sim 0.16$, the Einasto profile gives
better fit to the inner parts of halos over a broad mass range
\citep{Navarro2010, springel08_aqu, Gao2012}. The lensing properties of 
such a profile has also been investigated \citep[e.g.]{RM2012b,Dhar2010}.

\begin{figure}
 \includegraphics[width=0.5\textwidth]{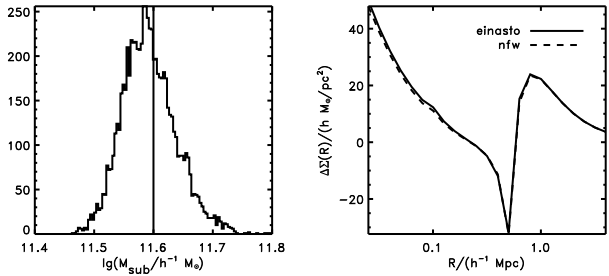}
 \caption{NFW fit to Einasto profile: the input subhalo lensing signal
is generated with Einasto haloes while the fitting assumes a NFW
profile.  Left panel shows the marginalized 1-d distribution of
subhalo mass, with the vertical line indicating the input mass.
In the right panel, solid line shows the input lensing signal and the 
dashed line shows the lensing signal produced by the best-fit  
NFW profile.}
\label{fig:einasto}
\end{figure}

Different density profile in the inner region can give different lensing signal. 
In galaxy-galaxy lensing, measurements usually cannot go into very
inner regions of halos. Thus, even the inner profiles differ from 
the NFW form, it can still give a good fit to the overall lensing
signal and recover the halo mass \citep{Mandelbaum2008}.  
We have also tested this effect by assuming that the input subhalo 
lensing signal is given by an Einasto profile, while assuming a NFW 
profile in  the fitting. Specifically , the Einasto halo we used has
$\alpha=0.16$, and $r_{-2}=r_{\rm s,sub}$, where $r_{\rm s,sub}$ is
the NFW scale radius of a halo with the same mass. 
We set the halo-centric distance to be $0.5 \mpch$  and the 
host halo mass to be $10^{14}\ms$.  The fitting result with 
the LEV2 noise model is shown in Fig. \ref{fig:einasto}. 
It is clear that the assumption of a NFW profile to fit Einasto 
haloes leads to negligible difference.

\section{Summary}
\label{sec:summary}

In this paper, we have investigated the feasibility of constraining
the masses and density profiles of dark matter subhaloes associated
with satellite galaxies using galaxy-galaxy lensing.  With the use of
a group catalog constructed from a large redshift survey, such as the
SDSS, we can effectively distinguish central and satellite
galaxies. Therefore we can select satellite galaxies according to
their host halo mass, halo-centric distance and stellar mass. By
stacking the galaxy-galaxy lensing signal of satellite galaxies with
similar properties, we can then study both the host halo and subhalo
density profiles. In this paper, we have used the SDSS group catalog
constructed by Yang et al. (2007) to predict the galaxy-galaxy lensing
signal for satellite galaxies residing in different host haloes and
located at different halo-centric distances.  To examine to what
extent such data can be used to infer the properties of host and
sub-halo, we have considered two different noise levels, LEV1 and
LEV2, that correspond to the levels of measurement noise expected from
a current generation galaxy survey, such as SDSS, and from a next
generation galaxy survey, such as LSST, respectively. Using the MCMC
method, we investigated how well LEV1 and LEV2-type data can constrain
the mass and density profile for subhaloes and host haloes
simultaneously. For satellite galaxies in massive groups, with host
halo masses in the range of $[10^{14},5\times10^{14}]\ms$, the density
profile of the host halo can be well constrained for both LEV1 and
LEV2 noise levels. However, the data quality that is achievable with
current LEV1-type surveys is insufficient to put any meaningful
constraints on the subhalo properties. In the case of a LEV2-type
survey, on the other hand, the galaxy-galaxy lensing data is predicted
to be of sufficient quality that one can put tight constraints on the
average subhalo mass, with a $1\sigma$ confidence of about $0.2$ dex.
Even the amplitude and characteristic scale of the subhalo density
profiles can be constrained, albeit with still relatively large
uncertainties.  We also demonstrate that, with LEV2-type surveys, it is
even possible to probe subhaloes in group-sized host haloes with masses
as low as $10^{13}\ms$.

We also discussed some potential systematics that result in biased
estimates. One of these arises from the fact that one stacks the
signal from satellite galaxies that span a significant range in
properties of host halo and subhalo.  Since the lensing signal does
not scale linearly with model parameters, the best fit of the mean
value of the parameters can be biased relative to the underlying
values of the stacked sample. We have shown that such bias can be
reduced by dividing the satellite sample into finer stellar mass bins
and using parameterized forms for the scaling relations between
satellite and subhalo properties.  Another bias arises from the
presence of interlopers in the group catalog (i.e., from galaxies that
have erroneously been assigned to a group due to projection effects).
This implies that some of the galaxies identified as satellites in the
group catalog are actually centrals of other (typically low-mass)
haloes. Our test using a mock SDSS group catalog shows that about
$10\%$ of the satellites are such interlopers.  Such a contamination
can bias the inferred subhalo mass high by $\sim 50\%$.  To overcome
the bias effect, we propose to include the contamination fraction as a
free parameter in the model fitting. Our test shows that the bias in
the subhalo mass due to the contamination can be effectively removed
at the expense of enlarged uncertainties. This uncertainty, in turn,
can be reduced by using priors on the interloper fractions which 
can be obtained from realistic mock galaxy redshift surveys.
We conclude that measurement of galaxy-galaxy lensing around
satellite galaxies with future surveys such as LSST holds great
promise for constraining the properties of dark matter substructure.

\section*{Acknowledgments}

LR is supported by China Postdoctoral Science Foundation, Grant
NO. 2011M500395.  Part of the computation was carried out on the SGI
Altix 330 system at the Department of Astronomy, Peking
University. HJM would like to acknowledge the support of 
NSF AST-1109354 and NSF AST-0908334.
ZHF is supported in part by the NSFC of China under grants
11033005 and 11173001. 

\bibliography{lensing}

\end{document}